\documentclass[10pt,twocolumn,tightenlines,floats,aps,amsmath,amssymb,prd,floatfix,longbibliography,superscriptaddress,notitlepage,nofootinbib]{revtex4-2}

\usepackage[utf8]{inputenc}
\usepackage[T1]{fontenc}
\usepackage{wasysym}
\usepackage{amsmath,amssymb,amsfonts,amsthm,mathrsfs}
\usepackage{graphicx,wrapfig}
\usepackage{enumerate}
\usepackage[table]{xcolor}
\definecolor{lightblue}{RGB}{51,102,204} % adjust values as desired
\definecolor{lightred}{RGB}{191,60,45} % adjust values as desired
\usepackage{physics}
\usepackage{mathtools}
\usepackage{dcolumn}
\usepackage{orcidlink}

\usepackage{hyperref}
\hypersetup{
	bookmarks=true,         % show bookmarks bar?
	unicode=false,          % non-Latin characters in Acrobat’s bookmarks
	pdftoolbar=true,        % show Acroba,t’s toolbar?
	pdfmenubar=true,        % show Acrobat’s menu?
	pdffitwindow=false,     % window fit to page when opened
	pdfstartview={FitH},    % fits the width of the page to the window
	pdftitle={Precision predictions of Starobinsky inflation with self-consistent Weyl-squared corrections},    % title
	pdfauthor={Mauricio Gamonal, Eugenio Bianchi},     % author
	pdfkeywords={inflation, cosmology, quantum gravity}, % list of keywords
	pdfnewwindow=true,      % links in new PDF window
	colorlinks=true,       % false: boxed links; true: colored links
	linkcolor=lightred,          % color of internal links (change box color with linkbordercolor)
	citecolor=lightblue,        % color of links to bibliography
	filecolor=cyan,         % color of file links
	urlcolor=lightblue        % color of external links
}

\newcommand{\ii}{\mathrm{i}}
\newcommand{\ee}{\mathrm{e}}

\newcommand{\outcomment}[1]{}

\begin{document}

\title{\Large Precision predictions of Starobinsky inflation\\[.2em]
with self-consistent Weyl-squared corrections}

\author{Eugenio Bianchi\,\orcidlink{0000-0001-7847-9929}}
\thanks{}
\email{ebianchi@psu.edu}
\author{Mauricio Gamonal\,\orcidlink{0000-0002-0677-4926}}
\thanks{}
\email{mgamonal@psu.edu}

\affiliation{Institute for Gravitation and the Cosmos, The Pennsylvania State University, University Park, Pennsylvania 16802, USA}\affiliation{Department of Physics, The Pennsylvania State University, University Park, Pennsylvania 16802, USA}

\date{\today}

\begin{abstract}
Starobinsky's $R+\alpha R^2$ inflation provides a compelling one-parameter inflationary model that is supported by current cosmological observations. However, at the same order in spacetime derivatives as the $R^2$ term, an effective theory of spacetime geometry must also include the Weyl-squared curvature invariant $W^2$. In this paper, we study the inflationary predictions of the gravitational theory with action of the form $R+\alpha R^2 - \beta W^2$, where the coupling constant $\alpha$ sets the scale of inflation, and  corrections due to the $W^2$ term are treated self-consistently via reduction of order in an expansion in the coupling constant $\beta$, at the linear order in $\beta/\alpha$. Cosmological perturbations are found to be described by an effective action with a nontrivial  speed of sound $c_{\textrm{s}}$ for scalar and $c_{\textrm{t}}$ for tensor modes, satisfying the relation $c_{\textrm{t}}/c_{\textrm{s}} \simeq 1+ \frac{\beta}{6\, \alpha}$ during the inflationary phase. Within this self-consistent framework, we compute several primordial observables up to the next-to-next-to-next-to leading order (N3LO). We find the tensor-to-scalar ratio $r \simeq 3(1-\frac{\beta}{6\alpha})(n_\textrm{s}-1)^2$, the tensor tilt $n_{\textrm{t}}\simeq-\frac{r}{8}$ and the running of the scalar tilt $\mathfrak{a}_{\textrm{s}}\simeq-\frac{1}{2}  (n_{\textrm{s}} - 1)^2$, all expressed in terms of the observed scalar tilt $n_{\textrm{s}}$.
We also provide the corresponding corrections up to N3LO, $\mathcal{O}((n_{\textrm{s}} - 1)^3)$.
\end{abstract}

\maketitle
% Add the custom footnote with a double dagger marker
\begingroup
\renewcommand{\thefootnote}{$\ddagger$} % Use double dagger in math mode
\footnotetext{Both authors contributed equally to this work.}
\endgroup

% Reset footnote counter to ensure text footnotes start at 1
\setcounter{footnote}{0}

\section{Introduction}
\label{sec:Introduction}
Cosmic inflation \cite{Brout1978, Starobinsky1979, Starobinsky1980,Mukhanov1981,Sato:1981ds,Sato:1981qmu,Guth1981,Linde1982,Albrecht1982,Guth1982,Hawking1982,Linde1983} provides a mechanism for the production of primordial curvature perturbations with a nearly scale-invariant power spectrum. Its characteristic power-law dependence is tightly constrained by observations of the cosmic microwave background (CMB) \cite{Calabrese:2013jyk,Planck:2018jri,KeckCollaboration2021,ACT:2020frw,ACT:2025tim}. In models of single-field inflation \cite{Martin:2013tda, Kallosh:2025ijd}, the inflationary phase of expansion is driven by a minimally coupled scalar field rolling down a nontrivial potential, which introduces multiple free parameters to be fitted against observational data \cite{Martin:2024qnn}. On the other hand, one of the oldest inflationary models,  the Starobinsky model \cite{Starobinsky1980}, only requires a single parameter $\alpha$, the coupling constant of a higher-curvature correction in the effective action $R + \alpha R^2$, without the introduction of any additional scalar field,
serving as a reference inflationary model in model comparison \cite{Martin:2010hh,Ballardini:2024ado,Sorensen:2024ezb}. The apparent simplicity of this model is even more remarkable in light of observational bounds on the tensor-to-scalar ratio $r_{0.05} < 0.036 $ \cite{Planck:2018jri,KeckCollaboration2021}, which disfavors monomial potentials for a scalar field while remaining consistent with the value predicted by the Starobinsky model, $r \simeq 3\, (n_\mathrm{s} - 1)^2 \approx 0.003$. However, at the same order as the $R^2$ term, other curvature-squared terms  appear in the effective action of quantum gravity \cite{tHooft:1974toh,Stelle1977,Stelle1978,Birrell:1982ix,Donoghue:1994dn,Burgess:2003jk,Buchbinder:2017lnd,Donoghue:2022eay}, and their coupling constants are to be measured before predictions can be extracted from the theory. Moreover, the introduction of a Weyl-squared term results in new difficulties due to tensor ghosts, which have to be carefully treated before one can extract physical predictions from the theory. In the context of cosmic inflation, different proposals for taking into account higher curvature terms have been explored 
\cite{Berkin:1991nb,Weinberg:2008,Weinberg:2009wa,Clunan:2009er,Deruelle:2010kf,Deruelle:2012xv,Myrzakulov:2014hca,DeLaurentis:2015fea,Baumann2015,Myung:2014jha,Myung:2015vya,Ivanov:2016hcm,Salvio:2017xul,Cano:2020oaa,Anselmi2020,Giare:2020plo,Kaczmarek:2021psy,Gialamas:2021enw,Salvio:2022mld,Koshelev:2022olc,Kubo:2022dlx,DeFelice:2023psw,Khodabakhshi:2024med,Kubo:2025jla}. 
In this paper we consider an approach that allows us to extract predictions for inflationary observables that are robust as they are independent of whether the theory admits a field-theoretical \cite{Modesto:2011kw,Koshelev:2016xqb,Salvio:2018crh,Anselmi:2018ibi,Donoghue:2021cza,Buoninfante:2023ryt,Buccio:2024hys} or a quantum geometry \cite{Ashtekar2021,Ashtekar:2011ni,Bianchi:2010zs,Gozzini:2019nbo,Gielen:2013kla,Dittrich:2021gww,Agullo:2023rqq} ultraviolet completion: We treat the Weyl-squared term as a self-consistent correction to Starobinsky inflation via \emph{reduction of order} \cite{Bhabha:1946zz,Simon:1990ic}.

We consider a gravitational theory with the metric $g_{\mu\nu}$ as the only dynamical variable. The effective action $S[g_{\mu\nu}]$ is local, invariant under diffeomorphisms, and it includes all combinations of quadratic curvature terms:
\begin{equation}
S[g_{\mu\nu}] =\frac{1}{16\pi G} \int \dd[4]{x} \, \sqrt{-g} \,\big(R + \alpha\,  R^2 - \beta \, W^2 \big)\, ,
\end{equation}
where $\alpha>0$ and $\beta>0$ are coupling constants, $R$ is the Ricci scalar, and $W^2 \equiv W_{\mu\nu\rho\sigma}W^{\mu\nu\rho\sigma}$ is the Weyl-squared invariant. As the Friedman-Lema\^itre-Robertson-Walker (FLRW) metric is conformally flat, the $W^2$ term does not affect the dynamics of the homogeneous and isotropic background geometry. Therefore, exactly as in Starobinsky inflation, the theory predicts an inflationary phase with a typical scale uniquely determined by the coupling constant $\alpha$. We treat the Weyl-squared term as an effective correction to Starobinsky inflation by assuming that the coupling constant $\beta$ satisfies $\beta\ll\alpha$. The reduction of order method \cite{Bhabha:1946zz,Simon:1990ic} provides a self-consistent construction that selects physically relevant solutions and avoids the contribution of ghost modes. We present a systematic study of inflationary observables, including precision predictions  obtained via Green's function method for the fully expanded primordial power spectrum associated with a quasi-Bunch-Davies vacuum \cite{Auclair2022,Bianchi:2024qyp}, up to next-to-next-to-next-to-leading order (N3LO) corrections in the Hubble flow parameters. We find that the Weyl-squared term induces an effective speed of sound, both for tensor and scalars, with $c_{\mathrm{t}}/c_{\mathrm{s}}\simeq 1+\beta/(6\,\alpha)$, which characterizes the $W^2$ correction to inflationary observables.

The paper is organized as follows: In Sec.~\ref{sec:geoEFT}, we introduce the effective action for spacetime geometry, discussing the main assumptions and considerations for its use in primordial cosmology. In Sec.~\ref{sec:Reduction-Order}, we give a pedagogical description of reduction of order, a systematic procedure used to distinguish physically relevant solutions obtained from effective theories. In Sec.~\ref{sec:Background-Dynamics}, we determine the background dynamics of the theory in a homogeneous and isotropic spacetime, where a quasi-de Sitter epoch naturally emerges from the effective geometry induced by the higher curvature terms. In Sec.~\ref{sec:Cosmological-Perturbations}, we consider small metric perturbations around the inflationary background, and, by self-consistently applying the reduction of order method, we determine the physically relevant contributions to the quadratic action for scalar and tensor perturbations. In Sec.~\ref{sec:Inflationary-Observations}, we report the predictions of $R+\alpha\, R^2-\beta\, W^2$ inflation for the primordial power spectrum of scalar and tensor modes, including a relational set of observables in terms of the scalar tilt $n_{\mathrm{s}}-1$, which is summarized in Table~\ref{Tab:Results_Summaryv2}. Finally, in Sec.~\ref{sec:Discussion}, we discuss the main results of this work and its relevance for next-generation CMB and gravitational-wave observations.

We work in units with speed of light $c=1$, while keeping track of Newton's gravitational constant $G$ and Planck's constant $\hbar$. The metric signature is $(-+++)$. We denote derivatives with respect to cosmic time by $(\dot{\phantom{a}})$ and derivatives with respect to other variables by $(\phantom{1})^\prime$. 

\section{An effective theory of geometry}
\label{sec:geoEFT}

At curvature lengths $\ell$ much larger than the Planck length $\ell_P=\sqrt{G\hbar}\approx 10^{-35}\,\mathrm{m}$, a quantum theory of spacetime geometry can be described in terms of an effective action organized as an expansion in the number spacetime derivatives \cite{Donoghue:1994dn,Burgess:2003jk,Weinberg:2008}. The background structure is a four-dimensional manifold, symmetries of the action are the spacetime diffeomorphisms encoding the general covariance of the theory, and the dynamical variable is assumed to be a Lorentzian metric tensor $g_{\mu\nu}(x)$. The most general diffeomorphism-invariant action, up to four spacetime derivatives, is
\begin{align}
	\label{eq:sec1_action_original}
	&S[g_{\mu\nu}]= \int \dd[4]{x} \, \sqrt{-g}\,\Big( c_0 + c_2\, R \nonumber\\[.5em]
	&\qquad+ c_4\, R^2 +  \tilde{c}_{4}\, R_{\mu\nu} R^{\mu\nu} + \tilde{\tilde{c}}_{4 }\, R_{\mu\nu\rho\sigma} R^{\mu\nu\rho\sigma} + \cdots \Big) ,
\end{align}
where $g\equiv \det(g_{\mu\nu})$, $R_{\mu\nu\rho\sigma}$ is the Riemann curvature tensor, $R_{\mu\nu} = R^\alpha_{~\mu\alpha\nu}$ is the Ricci tensor, $R= R^\mu_{~\mu}$ is the curvature scalar, and $c_n$ are the coupling constants of terms that contain $n$ derivatives of the metric. The first two terms reproduce general relativity: The constants $c_0$ and $c_2$ have already been measured, and they are related to Newton's constant $G$ and the cosmological constant $\Lambda$ by the identification $c_0 =\frac{-2\Lambda}{16\pi G}$ and $c_{2}=\frac{1}{16\pi G}$. Using the identity $W^2=2R_{\mu\nu}R^{\mu\nu} -(2/3) R^2 + \mathcal{E}$, and dropping the Euler density $\mathcal{E} \equiv 4 R_{\mu\nu} R^{\mu\nu} - R^2 - R_{\mu\nu\rho\sigma}R^{\mu\nu\rho\sigma}$ because, by Gauss-Bonnet theorem, it is a total derivative and therefore it does not contribute to the local  dynamics, we can reduce the number of independent coupling constants. Therefore, only the $R^2$ and the $W^2$ terms remain at the four-derivative order, beyond general relativity. After a convenient renaming of the coupling constants, the effective action for spacetime geometry reads
\begin{equation}
	\label{eq:gEFT_full_action}
	S[g_{\mu\nu}] =\frac{1}{16\pi G}\! \int\! \dd[4]{x} \sqrt{-g} \,\big(-2\Lambda+  R + \alpha  R^2 - \beta W^2 \big).
\end{equation}
With these conventions, the coupling constants have dimensions $[G\hbar]=\textrm{length}^2$, $[\Lambda]=\textrm{length}^{-2}$, $[\alpha]=\textrm{length}^2$, and $[\beta] = \textrm{length}^2$. The effective action does not predict the value of these coupling constants, which have to be measured. Here we work under the assumption that there is a hierarchy of scales:
\begin{equation}
\label{eq:hierarchy}
    1\ll \frac{\beta}{G\hbar}\ll\frac{\alpha}{G\hbar}\ll \frac{1}{G\hbar \Lambda}\,.
\end{equation}
% \begin{equation}
%     G\hbar\ll \beta\ll\alpha\ll \Lambda^{-1}\,,
% \end{equation}
Cosmological observations are already constraining the physics of the primordial universe in the regime where $\alpha$ becomes relevant. If Starobinsky inflation is responsible for the inflationary expansion, then the length scale associated with $\alpha$ is expected to be about 5 orders of magnitude larger than the Planck scale, $\ell_{\mathrm{infl}}\sim \sqrt{\alpha}\sim 10^5\,\ell_P$. In this regime, the effective description \eqref{eq:gEFT_full_action} applies, and including both the $\alpha$ and the $\beta$ terms in the effective action is unavoidable.

The $\alpha R^2$ term introduces an additional scalar degree of freedom, commonly referred to as the \textit{scalaron}, together with a modification of the dynamics of the standard transverse tensor modes of general relativity. This scalar mode is typically stable and can be interpreted as the \textit{inflaton} $\varphi^{(E)}$, once one introduces a  field redefinition $g_{\mu\nu}\to (g_{\mu\nu}^{(E)}, \varphi^{(E)})$ to obtain the Einstein frame formulation of Starobinsky inflation  \cite{Maeda:1988ab,DeFelice2010,Ketov:2025nkr}. Therefore the $\alpha R^2$ term provides a natural mechanism that drives the inflationary phase, without the addition of an \emph{ad hoc} inflaton scalar field. Here we treat it geometrically, without the introduction of an \emph{ad hoc} redefinition.

The $\beta W^2$ term includes higher-order time derivatives that can lead to a pathological behavior if treated as fundamental, instead of effective. Specifically, the quadratic Lagrangian for perturbations around a cosmological background includes terms of the form $\mathcal{L} \sim \beta\, f(t)\, |\ddot{\Psi}(\vb{k}, t)|^2$, where $f(t)$ is a background-dependent function, and $\Psi(\vb{k}, t)$ denotes the scalar or the tensor perturbations, resulting in additional ghost degrees of freedom which pose a challenge to the consistency of the theory at the perturbative level \cite{DeFelice:2023psw}.

We adopt the assumption that $\beta\ll\alpha$ and that the physics of the $\beta W^2$ term introduces only corrections to Starobinsky inflation that are self-consistent in the limit $\beta\to 0$. This minimal assumption reflects the observational evidence that only three gravitational degrees of freedom are relevant at the inflationary scale: the two transverse-traceless tensor polarizations predicted by general relativity, and confirmed by the direct detection of gravitational waves \cite{LIGOScientific:2016aoc,LIGOScientific:2017zic,LIGOScientific:2020tif}, and a single scalar degree of freedom, the scalaron, that naturally provides a dynamical mechanism for the generation of the observed spectrum of primordial curvature perturbations \cite{Starobinsky1980,Mukhanov1981,Vilenkin1985}. Consequently, contributions induced by the $\beta W^2$ are treated as effective corrections, in a similar fashion to the effective field theory approach advocated by Weinberg in \cite{Weinberg:2008}, with any higher time derivative reduced to obtain a self-consistent expansion that is analytic in the coupling constant $\beta$. This procedure ensures that the effective theory remains predictive and free of unphysical degrees of freedom at larger length scales.

In the next section, we present a pedagogical introduction to the method of \textit{reduction of order} that guarantees a self-consistent implementation of this prescription. 

\medskip

\section{Reduction of Order}
\label{sec:Reduction-Order}

Theories described by actions involving higher-than-first-order time derivatives generically lead to pathologies in the dynamics such as runaway solutions and the Ostrogradsky instability; see, e.g., \cite{Delhom:2022vae} for a pedagogical introduction, and \cite{Crisostomi:2017ugk} for a discussion in the context of higher derivative metric theories. This peculiar behavior is not merely of mathematical interest, but also relevant for a variety of physics problem. A notable example is in electrodynamics: the Lorentz–Abraham–Dirac equation \cite{Dirac:1938} describes the self-interaction of an accelerating charged particle and includes a third-order time derivative in its equation of motion, which leads to unphysical effects such as preacceleration and runaway solutions. These issues can be mitigated under reasonable physical assumptions, e.g., by treating the radiation reaction as a small perturbative correction to the Lorentz force. Such an approach leads to the Landau–Lifshitz equation \cite{Landau:1975}, a second-order differential equation that provides a consistent physically meaningful description. Remarkably, recent experimental efforts are starting to provide supporting evidence for the Landau–Lifshitz formulation \cite{Cole:2017zca, Nielsen:2021ppf}.

Using the Lorentz–Abraham–Dirac equation as an example, Bhabha \cite{Bhabha:1946zz} identified a key insight: imposing analyticity in the coupling constant associated with the higher-order time derivative corrections is necessary to distinguish physical solutions from unphysical ones. The application of these ideas in the context of gravity with higher-order curvature terms was also discussed by Simon in \cite{Simon:1990ic}, and a covariant formulation in the context of linearized semiclassical gravity was implemented in \cite{Frob:2013ht}. Here we follow a similar logic.

There are three relevant conditions that an effective theory has to satisfy for the self-consistent implementation of the reduction of order of higher derivative terms with coupling constant $\beta$:
\begin{enumerate}
	\item[$C_1$:] \textbf{Coupling constants and stability}.
        The solutions of the effective theory must be stable under small perturbations of the initial conditions. This condition restricts the range of the coupling constant $\beta$ to values that avoid unphysical runaway solutions.
	\item[$C_2$:] \textbf{Initial conditions and analyticity}. The initial conditions are restricted to the class that determines solutions that are analytic in an expansion around $\beta=0$, so that we can consistently consider the uncoupled limit.
	\item[$C_3$:] \textbf{Regime of validity}. The effective dynamics is defined by an asymptotic expansion in a dimensionless coupling constant $\lambda\propto \beta$ which is linear in $\beta$ and constructed from the characteristic scales of the problem, with the requirement  $\abs{\lambda}\ll 1$ for the leading order corrections to provide accurate predictions.
\end{enumerate}
To illustrate this procedure, we consider a simple toy model: a classical oscillator with mass $m_0>0$, frequency $\omega_0>0$, and with an extra acceleration term with coupling constant $\beta$ of unspecified sign and dimension $[\beta]= \textrm{time}^2$,  which is described by the action
\begin{equation}
	\label{eq:Action-Oscillator-Acceleration}
	S[x] =  \int \dd{t} m_0\qty[\frac{ \dot{x}(t)^2}{2} - \frac{\omega_0^2 x(t)^2 }{2} - \frac{\beta \ddot{x}(t)^2}{2}]. 
\end{equation}
The variational principle $\delta S = 0$ yields the equation of motion
\begin{equation}
	\label{eq:App_EoM_toymodel1}
\ddot{x}(t) + \omega_0^2 \, x(t) + \beta\, \ddddot{x}(t) = 0,
\end{equation}
which admits exact solutions of the general form
\begin{equation}
	x(t) = A_+\, \ee^{\ii \Lambda_{+} t} + B_+\, \ee^{-\ii \Lambda_{+} t} + A_-\, \ee^{\ii \Lambda_{-} t} + B_-\, \ee^{-\ii \Lambda_{-} t},
    \label{eq:exact-x}
\end{equation}
where $A_+$, $B_+$, $A_-$, and $A_-$ are integration constants to be determined by specifying the initial conditions $x(t_0)$, $\dot{x}(t_0)$, $\ddot{x}(t_0)$, and $\dddot{x}(t_0)$. The characteristic frequencies are
\begin{equation}
	\Lambda_{\pm} = \sqrt{\frac{1\pm \sqrt{1-4\beta \omega_0^2}}{2\beta\omega_0^2}}\;\omega_0 \,.
\end{equation}
From this expression we see that the relevant dimensionless coupling constant of the problem is $\lambda = \beta\, \omega_0^2$. We can now analyze the three necessary conditions for the validity of the effective theory, starting from the general exact solution \eqref{eq:exact-x}. 

\begin{figure}[t]
	\includegraphics[width=\linewidth]{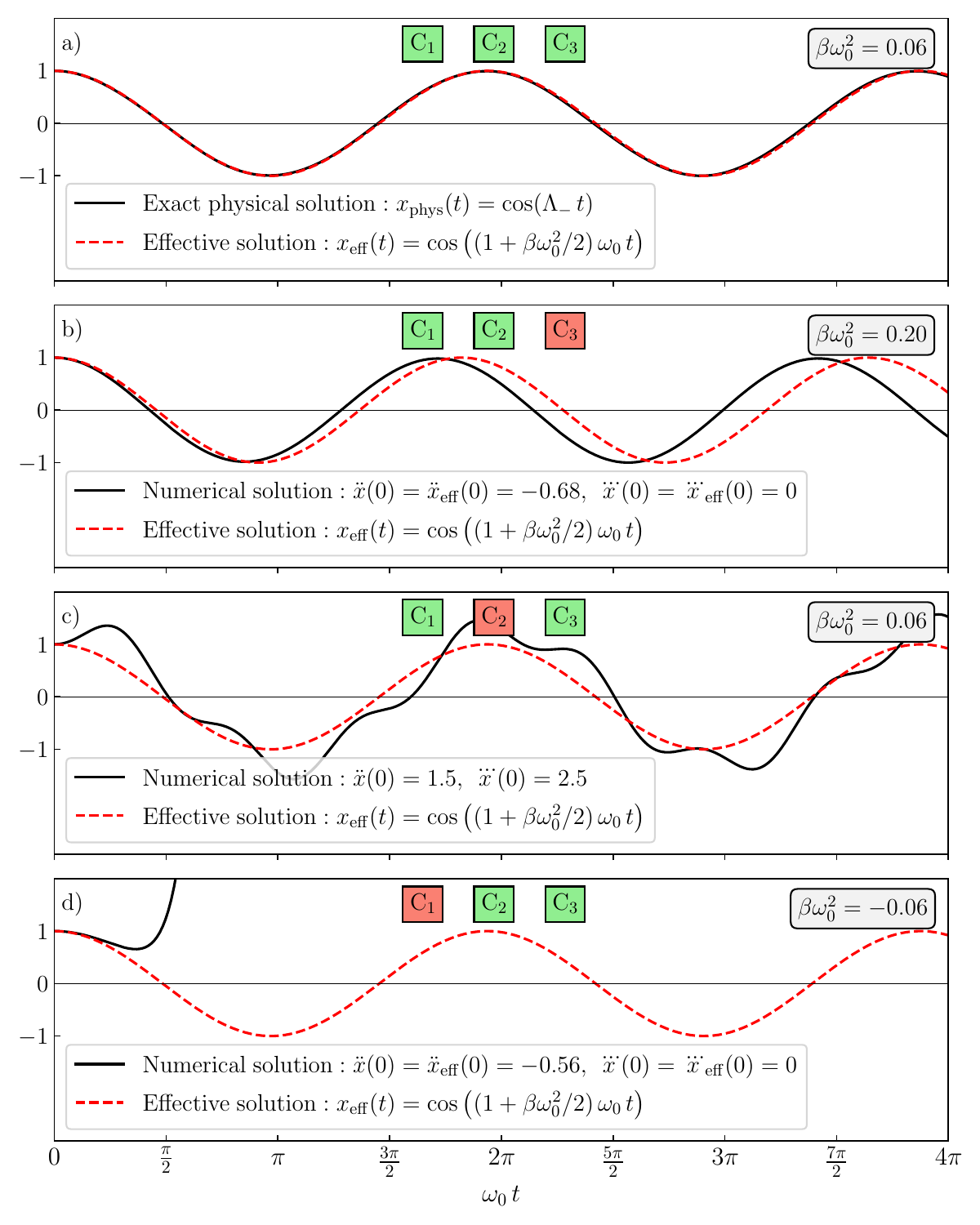}\quad
	\caption{A concrete example based on the action \eqref{eq:Action-Oscillator-Acceleration} which illustrates the three necessary conditions  for a self-consistent solution from reduction of order. The numerical solution refers to \eqref{eq:App_EoM_toymodel1}. Panel a): All three conditions are satisfied and the exact physical solution \eqref{eq:Reduction-Physical-Exact} matches almost perfectly the effective solution \eqref{eq:Reduction-Effective-Solution}.~Panel b):~Condition $C_3$ is not satisfied, as $\beta \omega_0^2\approx 0.2$, which goes slightly beyond the regime of validity of the lowest order truncation. Panel c): Condition $C_2$ is not satisfied, as initial conditions are chosen arbitrarily, and both frequencies $\Lambda_{+}$ and $\Lambda_{-}$ play a role in the dynamics, introducing a nonanalytic dependence on $\beta$. Panel d): Condition $C_1$ is not satisfied, as $\beta<0$ corresponds to an instability, and the characteristic runaway solution quickly becomes dominant.}
	\label{fig:Reduction-Order}
\end{figure}

The condition $C_1$ requires that the frequencies $\Lambda_\pm$ are real to avoid unphysical runaway exponentials. This condition requires 
\begin{equation}
    0\leq \beta \leq \frac{1}{4\,\omega_0^2}\,,
    \label{eq:beta-condit}
\end{equation}
which imposes a bound on the range of the coupling constant $\beta$. 

The condition $C_2$ requires analyticity in $\beta$ of the solution is determined by effective initial conditions. At the leading order in $\beta\, \omega_0^2$, the frequencies of the exact solution \eqref{eq:exact-x} are
\begin{align}
	\Lambda_{+} &= \frac{\omega_0}{\sqrt{\beta\, \omega_0^2}} + \order{\sqrt{\beta\, \omega_0^2}}\,, \label{eq:Lambda+} \\[.5em]
	\Lambda_{-} &= \qty(1+\frac{1}{2} \beta\, \omega_0^2)\,\omega_0\,   + \order{\beta^2\, \omega_0^4}. \label{eq:Lambda-}
\end{align}
Clearly, exact solutions that include $\Lambda_{+}$ terms are not analytic in $\beta$, which requires that the initial conditions select solutions with $A_+=0$ and $B_+=0$. In this way, we reduce the four-dimensional space of solutions to the ordinary two-dimensional space of solutions determined by the initial conditions $x(t_0)$ and $\dot{x}(t_0)$. For instance, the initial conditions $x(0)=1$, $\dot{x}(0)=0$ are completed by the self-consistent higher-order conditions $\ddot{x}(0)= -\Lambda_{-}^2$ and $\dddot{x}(0) = 0$, which determine the exact physical solution of the equation of motion \eqref{eq:App_EoM_toymodel1}:
\begin{equation}
	\label{eq:Reduction-Physical-Exact}
	x_{\mathrm{phys}} (t) = \cos(\Lambda_{-}\, t).
\end{equation}
This expression is automatically analytic in  $\beta$ and represents a well-defined exact physical solution of \eqref{eq:App_EoM_toymodel1}.

Finally, the condition $C_3$ requires that $|\beta \omega_0^2|\ll 1$, so that the solution can be obtained as an asymptotic series. In fact, except for simple toy models, exact physical solutions are often impossible to obtain, while an effective solution as a series in $\beta$ is immediate.

The reduced action $S_{\mathrm{red}}[x]$ is obtained by substituting iteratively the solution $\ddot{x}(t) =- \omega_0^2 \, x(t) - \beta\, \ddddot{x}(t)$ of \eqref{eq:App_EoM_toymodel1} into the action \eqref{eq:Action-Oscillator-Acceleration}, and then truncating the result at a finite order in $\beta$. This procedure leads to a self-consistent reduction of order and an expansion in powers of $\beta$. In this paper we consider only the first-order corrections in $\beta$ and neglect all terms of order $\order{\beta^2}$. Therefore, to obtain the first-order correction, we can simply substitute the zero-order equation of motion  $\ddot{x} (t)= -\omega_0^2 \, x(t)$ into the action \eqref{eq:Effective-Action-Oscillator-Acceleration}.  The reduced action at first order in $\beta$ is then
\begin{equation}
	\label{eq:Effective-Action-Oscillator-Acceleration}
	S_{\mathrm{red}}[x] =  \int \dd{t}  m_0\qty[\frac{ \dot{x}(t)^2}{2} - \frac{\omega_0^2(1+\beta \omega_0^2)}{2}  x(t)^2 ],
\end{equation}
which reduces to a simple harmonic oscillator with an effective frequency $\omega_{\mathrm{eff}}^2 = \omega_0^2\,(1+\beta\,\omega_0^2)$. The effective solution, with initial conditions $x(0)=1$ and $\dot{x}(0)=0$, is given by
\begin{equation}
\label{eq:Reduction-Effective-Solution}
x_{\mathrm{eff}}(t) = \cos\big( \omega_{\mathrm{eff}} t \big) = \cos\big((1+\beta\,\omega_0^2/2)\,\omega_0 t\big).
\end{equation} 
This effective solution accurately captures the exact physical solution \eqref{eq:Reduction-Physical-Exact} for which the conditions $C_1$, $C_2$ and $C_3$ hold, since $ \Lambda_{-} = \omega_{\mathrm{eff}} + \order{\beta^2\,\omega_0^4}$, as seen in \eqref{eq:Lambda-}. Note also that the effective solution provides a criterion to find approximate initial conditions for the exact solution. For instance, if we consider $x(0)=x_{\mathrm{eff}}(0)=1$, $\dot{x}(0)=\dot{x}_{\mathrm{eff}}(0)=0$, then the approximate solution implies $\ddot{x}(0)=\ddot{x}_{\mathrm{eff}}(0)= -\omega_{\mathrm{eff}}^2$ and $\dddot{x}(0) = \dddot{x}_{\mathrm{eff}}(0)= 0$, which determines the exact solution of \eqref{eq:App_EoM_toymodel1} of the form,
\begin{equation}
	\label{eq:Reduction-Unphysical}
x_{\mathrm{exact}} (t) = \frac{  (\omega_{\mathrm{eff}}^2 - \Lambda_{-}^2  ) \cos( \Lambda_{+} t ) + (\Lambda_{+}^2 - \omega_{\mathrm{eff}}^2  ) \cos(\Lambda_{-} t )   }{ \Lambda_{+}^2  - \Lambda_{-}^2}\,.
\end{equation}
This exact solution coincides with the physical solution \eqref{eq:Reduction-Physical-Exact} only when $\omega_{\mathrm{eff}} = \Lambda_{-}$, an equality that can be reached only as an asymptotic expansion in $\beta\,\omega_0^2$. This example illustrates the limitations of blindly solving the equations of motion with higher-order time derivatives, in contrast to the effective dynamics associated with physical solutions, as shown in Fig.~\ref{fig:Reduction-Order}: In Panel a) we show the comparison between the effective solution and the exact physical solution \eqref{eq:Reduction-Physical-Exact} when all three conditions $C_1$, $C_2$ and $C_3$ hold. In panel b) the assumption $C_3$ is relaxed; as the dimensionless coupling $\beta \omega_0^2$ is no longer small, the effective solution requires $\order{\beta^2\,\omega_0^4}$ corrections to better capture the dynamics; In panel c) the assumption $C_2$ is relaxed; an arbitrary choice of initial conditions shows how both frequencies, $\Lambda_{-}$ and $\Lambda_{+}$, contribute to the behavior of the exact solution \eqref{eq:Reduction-Unphysical}, which remains oscillatory as the condition \eqref{eq:beta-condit} is satisfied. In panel d) the assumption $C_1$ is relaxed; the coupling constant $\beta<0$ results in an unphysical runaway solution.

This simple toy model illustrates clearly the three conditions that arise also in more involved situations. In particular, the case of a time-dependent mass $m(t)$ and frequency $\omega(t)$ is closely related to the study of primordial perturbations. We will see in the next sections how the above conditions ensure well-behaved and physically relevant primordial perturbations after applying the reduction-of-order method to the contributions associated with $\beta W^2$. Condition $C_1$ constrains the allowed values of the coupling $\beta$, as previously identified in the literature, e.g., $\beta \geq 0$ is required to avoid tachyonic instabilities \cite{DeFelice:2023psw}, and a more stringent bound $0 \leq \beta \leq 6\,\alpha$ was found in \cite{Anselmi2020}. Condition $C_2$ guarantees that all physical observables admit a smooth expansion in $\beta$, thereby allowing one to consistently capture the higher-derivative corrections that continuously connect to the $R+\alpha R^2$ limit as $\beta \to 0$. Under condition $C_3$ and the hierarchy \eqref{eq:hierarchy}, the inflationary predictions derived below remain valid in the regime $\beta / \alpha \ll 1$.

%In inflationary models with $\beta\, W^2$ contributions, certain restrictions have been identified in the literature, e.g., $\beta\geq 0$ required to avoid tachyonic instabilities \cite{DeFelice:2023psw}, and a more stringent bound $0\leq \beta\leq 6\, \alpha$ was identified in \cite{Anselmi2020}.

%%%%%%%%

\section{The inflationary background in the geometric framework}
\label{sec:Background-Dynamics}

\begin{figure}[t]
	\centering	\includegraphics[width=\linewidth]{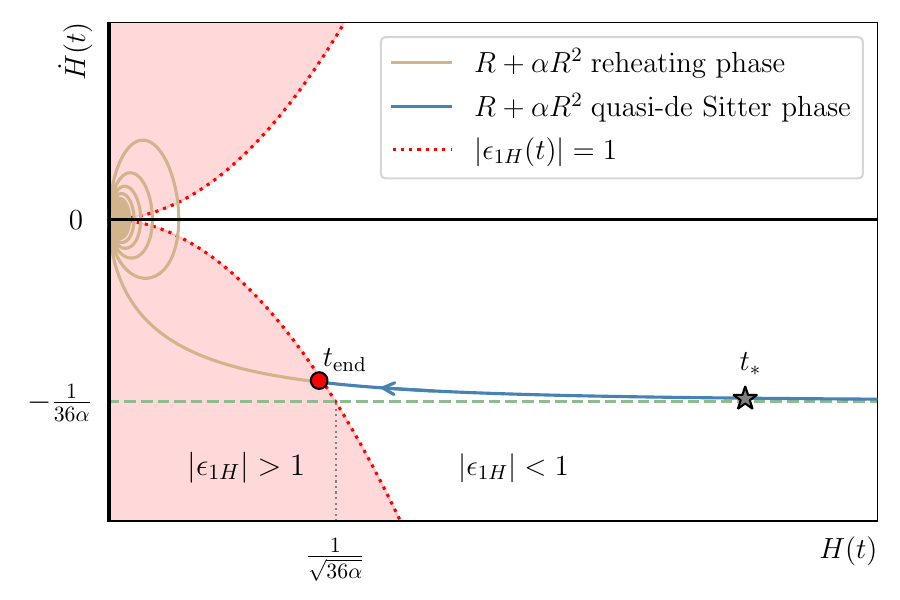}
	\centering
	\caption{Inflationary solution obtained from the modified Friedmann equation \eqref{eq:Friedmann_Modified}. In the geometric framework, a quasi--de Sitter inflationary phase (solid blue) arises naturally from the nontrivial dynamics of the Starobinsky action, followed by an oscillatory period of reheating (solid tan). The red dot indicates the end of the inflationary phase, defined as  $\ddot{a}(t_{\mathrm{end}})=0$ or equivalently by $\epsilon_{1H}(t_{\mathrm{end}})=1$, while the gray star denotes a typical freezing time used to define a pivot scale. At leading order, we have $\dot{H}\approx-1/36\alpha$, as shown in the green dashed line. The dotted gray line indicates the characteristic scale at the end of inflation, $H_{\mathrm{end}}^2 \approx 1/36\,\alpha$. }
	\label{fig:starobinsky}
\end{figure}

We consider a background spacetime geometry described by a spatially flat Friedman-Lema\^itre-Robertson-Walker (FLRW) metric,
\begin{align}
	\label{eq:FLRW-Metric}
	\bar{g}_{\mu\nu} \dd{x}^\mu \dd{x}^\nu &=- \dd{t}^2 + a(t)^2 \delta_{ij} \dd{x}^i \dd{x}^j \nonumber\\
	&= a(\eta)^2 \, (-\dd{\eta}^2 + \delta_{ij} \dd{x}^i \dd{x}^j)\, ,
\end{align}
with conformal time $\eta$ defined in terms of the cosmic time $t$ by $\dd{t} = a(\eta) \dd{\eta}$ as usual and $a(t)$ the scale factor. An inflationary phase of quasi--de Sitter expansion is defined as an epoch where the Hubble rate $H(t)\equiv \dot{a}(t)/a(t)$ varies slowly. To describe this phase, it is useful to introduce dimensionless Hubble flow parameters:
\begin{align}
	\epsilon_{1H}(t)\, &\equiv-\frac{H( t)}{H(t)^2}\,,\\[.5em]
\epsilon_{nH}(t)  \,&\equiv - \frac{\dot{\epsilon}_{n-1 \, H}(t)}{H(t)\,\epsilon_{n-1\, H }(t)} \, .
\end{align}
Moreover, because of the separation of scales, the cosmological constant $\Lambda$ can be neglected in the primordial universe and the gravitational action \eqref{eq:gEFT_full_action} reduces to
\begin{equation}
	\label{eq:ActionR+R2+W2}
	S[g_{\mu\nu}] =\frac{1}{16\pi G} \int \dd[4]{x} \, \sqrt{-g} \,\big( R + \alpha  R^2 - \beta W^2 \big). 
\end{equation}
We neglect also the Standard Model matter fields as their contribution becomes important only after the end of inflation, during the reheating phase. Therefore, the variational principle for the action \eqref{eq:ActionR+R2+W2} results in field equations with higher curvature terms,
\begin{equation}
	G_{\mu\nu}\,+\,\alpha\,\mathcal{H}_{\mu\nu}\,-4\,\beta\, \mathcal{B}_{\mu\nu}=\,0\,,
	\label{eq:G+H}
\end{equation}
in vacuum ($T_{\mu\nu}=0$) and with the covariantly conserved tensors  ($\nabla_\mu G^{\mu\nu}=0$, $\nabla_\mu \mathcal{H}^{\mu\nu}=0$, and $\nabla_\mu \mathcal{B}^{\mu\nu}=0$) defined by
\begin{align}
	G_{\mu\nu}&\;=\frac{1}{\sqrt{-g}}\frac{\delta}{\delta g^{\mu\nu}}\int \dd[4]{x} \sqrt{-g} \, R\,,\\[.5em]
	\mathcal{H}_{\mu\nu}&\;=\frac{1}{\sqrt{-g}}\frac{\delta}{\delta g^{\mu\nu}}\int \dd[4]{x} \sqrt{-g} \, R^2\,, \label{eq:Tensor-H}\\[.5em]
	\mathcal{B}^{\mu\nu} &\;=\frac{1}{4\,\sqrt{-g}}\frac{\delta}{\delta g^{\mu\nu}}\int \dd[4]{x} \sqrt{-g} \, W^{\alpha\beta\rho\sigma}W_{\alpha\beta\rho\sigma}\,.
\end{align}
Their components are given by
\begin{align}
	G_{\mu\nu}&\;= R_{\mu\nu}-\tfrac{1}{2} R\, g_{\mu\nu}\,,\label{eq:G-def}\\[.5em]
	\mathcal{H}_{\mu\nu}&\;= 2\big(R\,G_{\mu\nu}-\nabla_{(\mu} \nabla_{\nu )}R+ (\square R+\tfrac{1}{4} R^2)g_{\mu\nu}\big) \,,\label{eq:H-def}  \\[.5em]
		\mathcal{B}^{\mu\nu} &\;= \nabla_\alpha \nabla_\beta W^{\mu\alpha\nu\beta} + \tfrac{1}{2} R_{\alpha\beta} W^{\mu\alpha\nu\beta} ,
\end{align}
where $\square=g^{\mu\nu}\nabla_\mu\nabla_\nu$. The variation of the $\beta\, W^2$ term results in the Bach tensor $\mathcal{B}_{\mu\nu}$, which vanishes on conformally flat spacetimes, such as the FLRW background \eqref{eq:FLRW-Metric}. Therefore the term $\beta\, W^2$ in the effective action does not affect the background dynamics which is the same as the one described by the Einstein-Starobinsky equations $G_{\mu\nu} + \alpha\, \mathcal{H}_{\mu\nu} = 0$. When evaluated on the FLRW metric \eqref{eq:FLRW-Metric}, we obtain a modified Friedmann equation:
\begin{equation}
	\label{eq:Friedmann_Modified}
	H(t)^2 - 36\, \alpha\, H(t)^4\, \epsilon_{1H}(t)\, \big(1-\tfrac{1}{2} \epsilon_{1H}(t) - \tfrac{1}{3}\epsilon_{2H}(t) \big) = 0\,.
\end{equation}
A direct consequence of including the higher curvature terms in \eqref{eq:ActionR+R2+W2}, combined with the homogeneity and isotropy of the background geometry, is the natural emergence of a quasi--de Sitter phase, with $\dot{H}\approx -1/36\alpha$  \cite{Ruzmaikina1969}, as shown in Fig.~\ref{fig:starobinsky}. This is the geometric formulation of Starobinsky inflation \cite{Starobinsky1980,Starobinsky:1983zz,Vilenkin1985}, one of the oldest inflationary models, and initially motivated by quantum-gravity arguments on the renormalization of the energy-momentum tensor. In this geometric framework (also known as the Jordan frame), spacetime is described by the physical metric $g_{\mu\nu}$. Often a conformal transformation is introduced to cast the theory as a scalar field $\varphi^{(E)}$ minimally coupled to an auxiliary metric $g_{\mu\nu}^{(E)}$ (Einstein frame).  Although this field redefinition, $g_{\mu\nu}\to (g_{\mu\nu}^{(E)},\varphi^{(E)})$, is widely used in the literature as it can simplify some computations without modifying physical predictions---as long as observables are properly mapped in the new variables \cite{Capozziello1997,Faraoni1999,Karam2017,Ketov:2024klm,Toyama:2024ugg}---it is important to highlight that observations of the reheating era may distinguish between a minimal coupling of Standard Model fields to the metric $g_{\mu\nu}$ versus minimal coupling to the auxiliary metric $g_{\mu\nu}^{(E)}$ \cite{Appleby:2009uf,Kannike:2015apa,vandeBruck:2016leo,Nandi:2019xlj,Nandi:2019xve,Dorsch:2024nan,Ketov:2025nkr}. In this paper we work exclusively in the geometric framework (Jordan frame).

Another notable feature of the geometric framework is that the modified Friedmann equation \eqref{eq:Friedmann_Modified} admits a systematic expansion of any function of geometric quantities uniquely in terms of $\epsilon_{1H}(t)$. In particular, for $H(t)$ we have \cite{Bianchi:2024qyp},
\begin{align}
	\label{eq:H-to-eps}
	H(t) &= \frac{1}{6\sqrt{\alpha\, \epsilon_{1H}(t)}} \left[  1 - \frac{1}{12} \epsilon_{1H}(t)+ \frac{19}{288} \epsilon_{1H}(t)^2  \right.  \nonumber \\
	& \;\;\;\left. - \frac{373}{3456} \epsilon_{1H}(t)^3 + \frac{44035}{165888} \epsilon_{1H}(t)^4 + \order{\epsilon^5} \right] .
\end{align}
Similarly, for $\epsilon_{2H}(t)$, $\epsilon_{3H}(t)$ and $\epsilon_{4H}(t)$, we find
\begin{align}
	\label{eq:Epsilon_Expansion_234}
	\epsilon_{2H}(t)&= -2 \epsilon_{1H}(t) + \frac{1}{3} \epsilon_{1H}(t)^2 - \frac{5}{9} \epsilon_{1H}(t)^3 \nonumber \\ &\quad + \frac{38}{27} \epsilon_{1H}(t)^4 + \order{\epsilon^5}\,, \nonumber \\
	\epsilon_{3H}(t) &= -2 \epsilon_{1H}(t) + \frac{2}{3} \epsilon_{1H}(t)^2 - \frac{5}{3} \epsilon_{1H}(t)^3 + \order{\epsilon^4} \,,\nonumber \\[.5em]
	\epsilon_{4H}(t) &= -2   \epsilon_{1H}(t)  +  \epsilon_{1H}(t)^2 + \order{\epsilon^3}\,.
\end{align}
The expression for $H(t)$ in \eqref{eq:H-to-eps}, combined with \eqref{eq:Epsilon_Expansion_234}, solves the Friedmann equation \eqref{eq:Friedmann_Modified}, up to $\order{\epsilon^4}$, representing excellent approximations during the quasi--de Sitter epoch. The Hubble flow expansion is also useful to express the number of e-foldings $N$ as an integral over $\epsilon_{1H}$. This quantity, the
amount of cosmological expansion from a given time $t$ until the end of inflation, is defined by $a(t_{\mathrm{end}})=\ee^{N}\,a(t)$  and can be expressed as
\begin{equation}
	\label{eq:Ns-via-epsilon}
	N \equiv \int_{t}^{t_{\textrm{end}}} H(t') \dd{t'} = - \int_{\epsilon_{1H}}^{1} \frac{\dd{\epsilon_{1H}'}}{ \epsilon_{1H}'\;\, \epsilon_{2H}\big(\epsilon_{1H}'\big)  } \,.
\end{equation}
Other geometric quantities can also be expressed in a similar way starting from the expansion of $H(t)$ in \eqref{eq:H-to-eps}. For instance, the dimensionless gap $\Delta$, defined as 
\begin{equation}
\label{eq:delta-definition}
\Delta \equiv \ln(\frac{a_{\mathrm{end}} H_{\mathrm{end}} }{a \, H}) = N - \ln(\frac{H}{H_{\mathrm{end}}}) \, 
\end{equation}
and relevant for the description of the reheating epoch, can also be expressed in terms of $\epsilon_{1H}$. Specific expressions for the quantities $N_\odot$ and $\Delta_\odot$ at the freezing time $t_\odot$ are given in Appendix \ref{app:Computation_N_odot}, and a detailed discussion is provided in Sec.~\ref{sec:Inflationary-Observations}.

\section{Cosmological Perturbations}
\label{sec:Cosmological-Perturbations}

As shown in Sec.~\ref{sec:Background-Dynamics}, the background dynamics is governed by the $R + \alpha\,R^2$ sector of the full action \eqref{eq:ActionR+R2+W2}. On the other hand, perturbations are sensitive to the $\beta W^2$ term. We treat the Weyl-squared term $\beta\,W^2$ self-consistently as a higher-order effective correction, without adding new degrees of freedom to the two transverse-traceless tensor polarizations and the scalaron of the $R + \alpha\,R^2$ theory. In this sense, this approach follows the same logic as Weinberg's effective field theory of inflation \cite{Weinberg:2008}, with the further requirement that the only relevant dynamical degrees of freedom are gravitational, without any additional matter field contribution.  Accordingly, the action naturally splits into a core sector and its effective correction,
\begin{equation}
S[g_{\mu\nu}] = S_0 [g_{\mu\nu}] + S_{\mathrm{eff}} [g_{\mu\nu}]\, ,
\end{equation}
where $S_0 [g_{\mu\nu}]$ is the usual Starobinsky action,
\begin{equation}
	\label{eq:Starobinsky_Action}
	S_0[g_{\mu\nu}] = \frac{1}{16\pi G} \int \dd[4]{x} \sqrt{-g} \, (R+\alpha R^2)\,,
\end{equation} 
and the most general combination of effective corrections at the quadratic order in curvature is encoded in the Weyl-squared action,
\begin{equation}
	\label{eq:Weyl_Action}
	S_{\mathrm{eff}}[g_{\mu\nu}] = \frac{1}{16\pi G} \int \dd[4]{x} \sqrt{-g} \, (-\beta\, W^{\mu\nu\rho\sigma} \, W_{\mu\nu\rho\sigma})\,,
\end{equation} 
treated via reduction of order. 
The dynamics of the scalar-vector-tensor (SVT) perturbations is determined by the tensor $F^{\mu\nu} [g_{\mu\nu}]$ obtained from the variation of the action \eqref{eq:Starobinsky_Action},
\begin{equation}
	F^{\mu\nu} [g_{\mu\nu}] \equiv\fdv{S_0}{g_{\mu\nu}}\,=\,-\frac{1}{16\pi G}\sqrt{-g}\,\big(G^{\mu\nu}+\alpha \,\mathcal{H^{\mu\nu}}\big)\,,
\end{equation}
where we used \eqref{eq:G-def} and \eqref{eq:H-def} and $\delta g_{\mu\nu}=-g_{\mu\alpha}g_{\nu\beta}\delta g^{\alpha\beta}$. We restrict to small perturbations $\delta  g_{\mu\nu}$ of the FLRW metric \eqref{eq:FLRW-Metric}, so that the spacetime geometry is given (exactly) by $g_{\mu\nu} = \bar{g}_{\mu\nu}+\delta  g_{\mu\nu}$, and 
\begin{equation*}
	F^{\mu\nu} [\bar{g}_{\mu\nu}+\delta g_{\mu\nu}] =\bar{F}_0^{\mu\nu}(t)\,+\,\bar{F}_1^{\mu\nu\rho\sigma}(t)\,\delta g_{\rho\sigma}+\mathcal{O}(\delta g^2)\,,
\end{equation*}
where $\bar{F}_1^{\mu\nu\rho\sigma}(t)=\eval{\delta^2 S_0/(\delta g_{\mu\nu} \, \delta g_{\rho\sigma})}_{g_{\mu\nu}=\bar{g}_{\mu\nu}}$. Note that, since the background metric satisfies the Friedman equation \eqref{eq:Friedmann_Modified}, we have $\bar{F}_0^{\mu\nu}=0$. Therefore, the action \eqref{eq:Starobinsky_Action}, at quadratic order in the perturbation, can be written as
\begin{equation}
	S_0[\bar{g}_{\mu\nu}+\delta g_{\mu\nu}] = \bar{S}_0^{(0)}(t) + S_0^{(2)} [\delta g_{\mu\nu}] ,
\end{equation}
with the quadratic action given by
\begin{equation}
S_0^{(2)} [\delta g_{\mu\nu}]  = \int\!\dd[4]{x}\tfrac{1}{2}\,\delta g_{\mu\nu}\bar{F}_1^{\mu\nu\rho\sigma}(t)\,\delta g_{\rho\sigma}+\mathcal{O}(\delta g^3)\,.
	\label{eq:S-pert}
\end{equation}
The homogeneity and isotropy of the background allows us to organize the perturbations into SVT representations of the $3d$ Euclidean group, $\delta g_{\mu\nu}=\delta g^{(\mathrm{s})}_{\mu\nu}+\delta g^{(\mathrm{v})}_{\mu\nu}+\delta g^{(\mathrm{t})}_{\mu\nu}$ which in the quadratic action decouple, as shown by working in the Fourier domain, where we use $\psi( \vb{k},t )$ to denote the Fourier transform of any SVT mode $\Psi(\vb{x},t)$, with
\begin{equation}
	\Psi(\vb{x},t) = \int \frac{\dd[3]{\vb{k}} }{(2\pi)^3}\,  \psi( \vb{k},t )\, \ee^{\ii  \vb{k}\cdot \vb{x}}.
	\label{eq:Fourier}
\end{equation}
The goal is to cast the quadratic contributions to the action, from both \eqref{eq:Weyl_Action} and \eqref{eq:S-pert}, for each SVT mode, in the form
\begin{widetext}
\begin{equation}
	\label{eq:Quadratic-Action-psi}
	S^{(2)}[\psi]= \! \int\!\! \dd{t} \!\!\int\!\! \frac{ \dd[3]{\vb{k}} }{(2\pi)^3}\, a(t)^3 Z_\psi(t) \Bigg[  \frac{|\dot{\psi}(\vb{k},t) |^2}{2} 
	-\frac{c_\psi(t)^2\,k^2}{a(t)^2}  \frac{\abs{\psi(\vb{k},t)}^2}{2} \Bigg],
\end{equation}
\end{widetext}
where the kinetic amplitude, $Z_\psi(t)$, and the speed of sound, $c_\psi(t)$, are two independent functions that do not depend on the mode $k$. We require also that they satisfy the no-ghost condition $Z_\psi(t)>0$, and the Laplacian stability condition $c_\psi(t)^2 >0$. This is exactly the general form of an effective theory of inflationary perturbations studied in \cite{Bianchi:2024qyp}, where we developed the formalism that allows us to provide explicit expressions for the fully expanded primordial power spectrum and other inflationary observables up to the next-to-next-to-next-to leading order (N3LO). 

\medskip

\subsection*{Scalar perturbations}
First, note that the Starobinsky action can be written in a form that allows us to recognize the gravitational scalar degree of freedom. If we define $\chi \equiv 1+2\,\alpha R$, then
\begin{equation}
	\label{eq:Relabed-Starobinsky-Action}
S_0 [g_{\mu\nu}] = \frac{1}{16\pi G} \int \dd[4]{x} \sqrt{-g} \, \qty(\chi \, R - \frac{1}{4\,\alpha} (\chi -1)^2).
\end{equation}
When $\chi$ is treated independently, and assuming small perturbations such that $\chi(\vb{x},t) = \bar{\chi}(t) + \delta\chi(\vb{x},t)$, the first variation of the action \eqref{eq:Relabed-Starobinsky-Action} gives
\begin{equation}
\square \chi(\vb{x},t) = \frac{1}{6\, \alpha}\, \chi (\vb{x},t)\, 
\end{equation}
where the emergence of a propagating scalar degree of freedom, the \emph{scalaron}, becomes evident, as typically found in $f(R)$ theories of gravity \cite{DeFelice2010}. For metric perturbations, we use the Arnowitt–Deser–Misner (ADM) variables 
\begin{align}
g_{\mu\nu}\dd x^\mu \dd{x}^\nu &= - N^2 \dd{t}^2  + h_{ij} (N^i \dd{t} + \dd{y}^i) (N^j \dd{t} + \dd{y}^j)\, , 
\end{align}
with $i,j=1,2,3$. Thus, the small metric perturbation $\delta g^{(\mathrm{s})}_{\mu\nu}$ can be expressed in terms of the lapse $N=1+\delta N$ and of the shift $N^i=0+\delta N^i$, with scalar perturbations $\delta N (\vb{k},t),$ and $C$, with $\delta N^i (\vb{k},t) = \ii \,k^i C(\vb{k},t)$, together with the $3d$ metric $h_{ij}=a(t)^2 \, ( \delta_{ij}+ \delta h_{ij}^{(\mathrm{s})} )$, with scalar perturbations $\mathcal{R}$ and $\mathcal{S}$, as
\begin{equation}
\delta h_{ij}^{(\mathrm{s})} (\vb{k},t) = 
-2\, \mathcal{R}(\vb{k},t) \, \delta_{ij} + k_i k_j \,\mathcal{S}(\vb{k},t)\, .
\end{equation}
In summary, we recognize five scalar perturbative quantities:
\begin{equation}
	\label{eq:Scalar-Perturbations}
\delta \chi(\vb{k},t),\; \delta N(\vb{k},t),\; C(\vb{k},t), \; \mathcal{S}(\vb{k},t), \; \mathcal{R}(\vb{k},t)\, .
\end{equation}
We work in the comoving gauge, setting $C(\vb{k},t)=0$ and $\mathcal{H}^{0}{}_i=0$ for the Starobinsky tensor \eqref{eq:Tensor-H}, which generalizes the comoving gauge $T^{0}{}_i=0$ for the energy-momentum in general relativity with matter. These conditions implicitly imply that $\delta \chi (\vb{k},t)= 0$. The diffeomorphism constraint $F^0{}_i\approx 0$ imposes
\begin{equation}
\delta N (\vb{k},t) = -\frac{2\, \bar{\chi}(t) \, \dot{\mathcal{R}}(\vb{k},t)}{2 H(t) \,\bar{\chi}(t)+ \dot{\bar{\chi}}(t) }\,,
\end{equation}
while in combination with the Hamiltonian constraint $F^0{}_0\approx 0$ we find
\begin{equation}
k^2\, \mathcal{S}(\vb{k},t) = \frac{ 3\, \epsilon_{1\chi}(t)^2 \, \dot{\mathcal{R}}(\vb{k},t) +4 \tfrac{k^2}{a(t)^2} \mathcal{R}(\vb{k},t) (1-\tfrac{\epsilon_{1\chi} (t)}{2})   }{  4 (1- \tfrac{\epsilon_{1\chi} (t)}{2}   )^2  } \, ,
\end{equation}
where we have introduced the background value for $\bar{\chi}(t)$,
\begin{align}
	\bar{\chi}(t) &\equiv 1+2\, \alpha\,  \bar{R}(t)\nonumber \\[.5em]
	&= 1+24\, \alpha \, H(t)^2\,  (1-\tfrac{1}{2}\epsilon_{1H}(t))\,, \label{eq:chi-bar}
\end{align}
together with its Hubble flow parameters
\begin{align}
	\epsilon_{1\chi}(t) &= - \frac{\dot{\bar{\chi}}(t)}{H(t)\,\bar{\chi}(t)}\,, \,	\epsilon_{2\chi}(t) = - \frac{\dot{\epsilon}_{1\chi}(t)}{H(t)\,\epsilon_{1\chi}(t)}\; , \;  \mathrm{etc}.
\end{align}
Therefore, we can express all five scalar perturbations \eqref{eq:Scalar-Perturbations} uniquely in terms of the curvature perturbation $ \mathcal{R}(\vb{k},t)$.  Note that, at first order, constant-$t$ spatial sections have a scalar curvature given by the $3d$ Ricci scalar ${}^{(3)\!}R=4a(t)^{-2}\,\delta^{ij}\partial_i \partial_j \mathcal{R}$. Substituting accordingly, we find that the quadratic contributions to the action $S_0$ in \eqref{eq:S-pert} for the curvature perturbation $\mathcal{R}(\vb{k},t$) take the form \eqref{eq:Quadratic-Action-psi}, with kinetic amplitude and speed of sound
\begin{align}
	Z^{(\alpha)}_{\textrm{s}}(t) \, &= \frac{3\bar{\chi}(t)}{16\pi G}  \qty(\frac{\epsilon_{1\chi}(t)}{1-\frac{1}{2}\epsilon_{1\chi}(t)})^2\,, 	\label{eq:Z-s-Starobinsky}\\[.5em]
	c^{(\alpha)}_{\textrm{s}}(t) \, &=\,1\,.
\end{align}
The stability of perturbations, also known as the no-ghost condition, requires $Z^{(\alpha)}_{\textrm{s}}(t)>0$, imposing the condition $\bar{\chi}(t)>0$, or $\epsilon_{1H}(t)<2 + \tfrac{1}{12\, \alpha\, H(t)^2}$, which is satisfied during the quasi--de Sitter phase of the background geometry derived from \eqref{eq:Friedmann_Modified}, and remains satisfied in the postinflationary universe. Moreover, from \eqref{eq:chi-bar},  we have $\bar{\chi}\to 1$ and $\epsilon_{1\chi}\to 0$ as $\alpha\to 0$, which shows that the perturbations are analytic in $\alpha$, i.e., $\lim_{\alpha \to 0} Z_{\mathrm{s}}^{(\alpha)}(t) = 0$, so the action reduces to 
standard general relativity without a propagating scalar degree of freedom.

The linearized first variation $F^{\mu\nu}$ provides the equation of motion for $ \mathcal{R}(\vb{k},t)$, which reads
\begin{equation}
	\label{eq:EoM-Scalars}
	\ddot{\mathcal{R}} (\vb{k},t)+ H(t) \, \qty(3-\epsilon_{1Zs}^{(\alpha)}(t)) \dot{\mathcal{R}} (\vb{k},t) + \frac{k^2}{a(t)^2} \, \mathcal{R}(\vb{k},t) =0,
\end{equation}
where 
\begin{equation}
\epsilon_{1Zs}^{(\alpha)} = - \frac{\dot{Z}_{\mathrm{s}}^{(\alpha)}}{H(t) Z^{(\alpha)}_{\mathrm{s}}} = \frac{\epsilon_{1\chi}(t) +2 \epsilon_{2\chi}(t) - \tfrac{1}{2} \epsilon_{1\chi}(t)^2 }{1-\tfrac{1}{2} \epsilon_{1\chi}(t)}. 
\end{equation}
To obtain the quadratic contributions to the effective action $S^{(2)}_{\mathrm{eff}}[\mathcal{R}]$, we substitute the previous expressions, gauge conditions and Hamiltonian constraints directly in \eqref{eq:Weyl_Action}. Therefore, the contributions to the full quadratic action of curvature perturbations $S^{(2)}[\mathcal{R}]$ can be expressed as
\begin{widetext}
	\begin{align}
		\label{eq:Quad-Action-Weyl-Scalar-Full}
		S^{(2)} [\mathcal{R}]	&
        = S_0^{(2)} [\mathcal{R}] +  S_{\mathrm{eff}}^{(2)} [\mathcal{R}] \nonumber \\[0.75em]
        &=  \! \int\!\! \dd{t} \!\!\int\!\! \frac{ \dd[3]{\vb{k}} }{(2\pi)^3} \, a(t)^3 Z_{\mathrm{s}}^{(\alpha)}(t) \Bigg( \frac{1}{2} |\dot{\mathcal{R}} (\vb{k},t)|^2 -  \frac{k^2}{a(t)^2}\frac{1}{2} |\mathcal{R}(\vb{k},t)|^2   \Bigg) \nonumber \\
        &\hspace{12em} -\beta\, \! \int\!\! \dd{t} \!\!\int\!\! \frac{ \dd[3]{\vb{k}} }{(2\pi)^3} \, a(t)^3 A_{\mathrm{s}}(t) \qty| B_{\mathrm{s}}(t)\, \dot{\mathcal{R}}(\vb{k},t) +C_{\mathrm{s}}(t) \, \Big[\frac{k^2}{a(t)^2} \mathcal{R}(\vb{k},t)+ \ddot{\mathcal{R}} (\vb{k},t) \Big] |^2  \nonumber \\[0.75em]
      &\hspace{-0.4em}\overset{\mathrm{R.O.}}{=} \! \int\!\! \dd{t} \!\!\int\!\! \frac{ \dd[3]{\vb{k}} }{(2\pi)^3} \, a(t)^3 \Bigg[Z_{\mathrm{s}}^{(\alpha)}(t) \Bigg( \frac{1}{2} |\dot{\mathcal{R}} (\vb{k},t)|^2 -  \frac{k^2}{a(t)^2}\frac{1}{2} |\mathcal{R}(\vb{k},t)|^2   \Bigg)\nonumber \\
      &\hspace{9em} - \beta  A_s(t) \Big(B_s(t) -C_s(t)\, H(t) \, \qty(3-\epsilon_{1Zs}^{(\alpha)}(t)) \Big)^2\, |\dot{\mathcal{R}} (\vb{k},t)|^2\Bigg]  \nonumber \\[0.75em]  &= \! \int\!\! \dd{t} \!\!\int\!\! \frac{ \dd[3]{\vb{k}} }{(2\pi)^3} \, a(t)^3 Z_{\mathrm{s}}(t) \Bigg( \frac{1}{2} |\dot{\mathcal{R}} (\vb{k},t)|^2 - c_{\mathrm{s}}(t)^2 \frac{k^2}{a(t)^2}\frac{1}{2} |\mathcal{R}(\vb{k},t)|^2   \Bigg),
	\end{align}
with
\begin{align*}
	A_{\mathrm{s}}(t) = \frac{1}{768\, \pi\, G } \frac{1}{(1-\tfrac{1}{2} \epsilon_{1\chi}(t))^6} \, , \;
	B_{\mathrm{s}}(t) = -6\, \epsilon_{1\chi}(t)^2\, 	H(t)\Big(1 - \tfrac{1}{2} \epsilon_{1\chi}(t) - 2\, \epsilon_{2\chi}(t)\Big)\, ,\;
	C_{\mathrm{s}}(t) = -6\, \epsilon_{1\chi}(t)^2\, \left(1 -\tfrac{1}{2} \epsilon_{1\chi}(t) \right)  \nonumber \, ,
\end{align*}
\end{widetext}
and where in the third equality we reintroduced the equation of motion \eqref{eq:EoM-Scalars} following the method of reduction of order (R.O.) discussed in Sec. \ref{sec:Reduction-Order}. The resulting action \eqref{eq:Quad-Action-Weyl-Scalar-Full} has effective kinetic amplitude and speed of sound given by
\begin{align}
\label{eq:Z-s-Weyl-Starobinsky}
Z_{\mathrm{s}} (t)&=  \frac{Z_{\mathrm{s}}^{(\alpha)}}{c_{\mathrm{s}}(t)^2} = \frac{3\,  \epsilon_{1\chi}(t)^2 }{16 \,\pi \,G} \, \Bigg( \frac{\bar{\chi}(t) - 2 \, \beta\, H(t)^2 \, \epsilon_{1\chi}(t)^2 }{(1-\tfrac{1}{2} \epsilon_{1\chi}(t))^2} \Bigg)\, , \nonumber  \\[0.75em]
\frac{1}{c_{\mathrm{s}}(t)^2} &= 1 - 2\, \beta \, \frac{H(t)^2  \, \epsilon_{1\chi}(t)^2}{\bar{\chi}(t)}\, ,
\end{align}
and now it takes the form  \eqref{eq:Quadratic-Action-psi}, for which a N3LO analysis of the inflationary predictions becomes available \cite{Bianchi:2024qyp}.  To illustrate the practicality of this procedure, we introduce new Hubble flow parameters associated with the kinetic amplitude $Z_\psi(t)$,
\begin{align}
	\epsilon_{1Z}(t)\, &\equiv-\frac{\dot{Z}_\psi( t)}{H(t) Z_\psi (t)}\,,\\
	\epsilon_{(n+1)Z}(t)\, &\equiv - \frac{ \dot{ \epsilon}_{nZ} (t) }{ H(t)  \,\epsilon_{nZ}(t)}\,,
\end{align}
and with the speed of sound $c_\psi(t)$,
\begin{align}
	\epsilon_{1c}(t) \,& \equiv -\frac{\dot{c}_\psi( t)}{H(t) c_\psi (t)}\,,\\
	\epsilon_{(n+1)c}(t) \,& \equiv - \frac{ \dot{ \epsilon}_{nc} (t) }{ H(t)  \,\epsilon_{nc}(t)}\,.
\end{align}
Hence, for the scalar modes, the expanded kinetic amplitude and speed of sound are given, up to N3LO, by
\begin{align}
Z_{\mathrm{s}}(t) &= \frac{\epsilon_{1H}(t)}{2 G \pi} \Big( 
		1 + \frac{5}{6} \epsilon_{1H}(t) - \frac{4}{9} \epsilon_{1H}(t)^2 
		- \frac{\beta}{3 \alpha} \, \epsilon_{1H}(t)^2 \nonumber\\
		& \quad - \frac{11}{27} \epsilon_{1H}(t)^3 
		+ \frac{13 \beta}{18 \alpha}\, \epsilon_{1H}(t)^3
		\Big), \nonumber \\[.5em]
        c_{\mathrm{s}} (t) &= 1 + \frac{\beta}{6 \alpha} \, \epsilon_{1H}(t)^2
		- \frac{\beta}{2 \alpha} \, \epsilon_{1H}(t)^3
		+ \frac{61 \beta}{72 \alpha}\, \epsilon_{1H}(t)^4 \, ,
\end{align}
and Hubble flow parameters 
\begin{align}
\epsilon_{1Z}^{(\mathrm{s})}(t) &= -2 \epsilon_{1H}(t) - \tfrac{4}{3} \epsilon_{1H}(t)^2 
		+ \tfrac{26}{9} \epsilon_{1H}(t)^3 
		+ \tfrac{4 \beta}{3 \alpha} \, \epsilon_{1H}(t)^3, \nonumber \\[.5em]
\;\epsilon_{2Z}^{(\mathrm{s})}(t) &= -2 \epsilon_{1H}(t) - \epsilon_{1H}(t)^2 
		+ \tfrac{19}{3} \epsilon_{1H}(t)^3 
		+ \tfrac{8 \beta}{3 \alpha}\, \epsilon_{1H}(t)^3,\nonumber  \\[.5em]
\;\epsilon_{3Z}^{(\mathrm{s})}(t) &= -2 \epsilon_{1H}(t) - \tfrac{2}{3} \epsilon_{1H}(t)^2 
		+ 17\, \epsilon_{1H}(t)^3 
		- \tfrac{16 \beta}{3 \alpha}\, \epsilon_{1H}(t)^3, \nonumber \\[1em]
\;\epsilon_{1c}^{(\mathrm{s})}(t) &= -\frac{2 \beta}{3 \alpha} \, \epsilon_{1H}(t)^3, \nonumber  \\[.5em]
\;\epsilon_{2c}^{(\mathrm{s})}(t) &= -6\,  \epsilon_{1H}(t) + \tfrac{31}{3} \epsilon_{1H}(t)^2 
		- \tfrac{40}{9} \epsilon_{1H}(t)^3 
		- \tfrac{4 \beta}{3 \alpha}\, \epsilon_{1H}(t)^3 , \nonumber \\[.5em]
\;\epsilon_{3c}^{(\mathrm{s})}(t) &= -2 \epsilon_{1H}(t) + \tfrac{34}{9} \epsilon_{1H}(t)^2 
		+ \tfrac{206}{81} \epsilon_{1H}(t)^3 
		- \tfrac{8 \beta}{9 \alpha}\, \epsilon_{1H}(t)^3 .
\end{align}

\subsection*{Vector perturbations}

We introduce the transverse vector fields $\delta N^i_T$  for the shift, and  $B_i(\vb{x},t)$, such that the $3d$ metric can be written as $h_{ij}=a(t)^2 \, ( \delta_{ij}+ \delta h_{ij}^{(\mathrm{v})} )$, with
\begin{equation}
\delta h_{ij}^{(\mathrm{v})} (\vb{k},t)  = \ii\,  (   
k_i B_j(\vb{x},t) + k_j B_i(\vb{k},t) ).
\end{equation}
Setting the same comoving gauge, and solving the transverse part of the diffeomorphism constraint, one finds that there is no propagating vector perturbation, as it already happens in conventional $f(R)$ theories \cite{DeFelice2010}. 

\medskip

\subsection*{Tensor perturbations}

For the transverse-traceless tensor perturbations, the $3d$ metric reads $h_{ij}=a(t)^2 \, ( \delta_{ij}+ \delta h_{ij}^{(\mathrm{t})} )$, with
\begin{equation}
	\delta h_{ij}^{(\mathrm{t})} (\vb{k},t)  =  e_{ij}^{(+)} (\vb{k}) \, \gamma_{+} (\vb{k},t) + e_{ij}^{(-)} (\vb{k})\, \gamma_{-} (\vb{k},t)\, ,
\end{equation}
where $e_{ij}^{(\pm)}(\vb{k})$ is the polarization tensor.  The equation of motion for tensor modes can be read from the linearized first variation $F^{\mu\nu}$, and is given by
\begin{equation}
	\label{eq:EoM-Tensors}
	\ddot{\gamma}_\sigma (\vb{k},t)+ H(t) \, \qty(3-\epsilon_{1Zt}^{(\alpha)}(t)) \dot{\gamma}_{\sigma} (\vb{k},t) + \frac{k^2}{a(t)^2} \,\gamma_\sigma (\vb{k},t) =0\, ,
\end{equation}
with $\sigma = \pm$, and
\begin{equation}
\epsilon_{1Zt}^{(\alpha)}(t) = \epsilon_{1\chi}(t) \, .
\end{equation}
Since the equation of motion \eqref{eq:EoM-Tensors} is the same for both polarizations, we denote $\gamma\equiv \gamma_\sigma$, and the quadratic contributions to the action $S_0$ in \eqref{eq:S-pert} can be computed by tracing over $e_{ij}^{(\pm)}(\vb{k})$ and summing over $\gamma_\sigma$. The resulting action $S^{(2)}[\gamma] $ now takes the form \eqref{eq:Quadratic-Action-psi}, with kinetic amplitude and speed of sound:
\begin{align}
	Z_{\textrm{t}}^{(\alpha)} (t)\,&= \frac{\bar{\chi}(t)}{64 \pi G}\,, \label{eq:Z-t-Starobinsky}\\[.5em]
	c_{\textrm{t}}^{(\alpha)}(t) \, &=\,1\,.
\end{align}
The condition for the stability of perturbations is also satisfied, as $\bar{\chi}(t)>0$ for the inflationary solution of the background equations. Furthermore, from \eqref{eq:chi-bar} it is clear that the perturbations are analytic in $\alpha$, i.e., $\lim_{\alpha \to 0} Z_t^{(\alpha)}(t) = 1/64\pi G$, and the action reduces to the one of gravitational waves propagating in standard general relativity.

Following the same procedure as for scalars, we can include the contribution of the $\beta W^2$ term and evaluate the effective action $S_{\mathrm{eff}}[\gamma]$ for tensor modes in  \eqref{eq:Weyl_Action}. Thus, using reduction of order (R.O.) and integration by parts (I.B.P.), the contributions to the full quadratic action of tensor perturbations $S^{(2)}[\gamma]$ can be expressed as
\begin{widetext}
\begin{align}
\label{eq:Quad-Action-Weyl-Tensor-Full}
S^{(2)} [\gamma]&= S_0^{(2)}[\gamma] + S_{\mathrm{eff}}^{(2)} [\gamma] \nonumber \\
&= \! \int\!\! \dd{t} \!\!\int\!\! \frac{ \dd[3]{\vb{k}} }{(2\pi)^3} \, a(t)^3 Z_{\mathrm{t}}^{(\alpha)}(t) \Bigg( \frac{1}{2} |\dot{\gamma} (\vb{k},t)|^2 - \frac{k^2}{a(t)^2}\frac{1}{2} |\gamma(\vb{k},t)|^2   \Bigg)  -\frac{\beta}{64 \pi G}\,  \! \int\!\! \dd{t} \!\!\int\!\! \frac{ \dd[3]{\vb{k}} }{(2\pi)^3} \, a(t)^3  \Bigg( \Big(H(t)^2- 4\frac{k^2}{a(t)^2} \Big)\,|\dot{\gamma}(\vb{k},t)|^2 \nonumber \\
&\qquad + 4\,\frac{k^2}{a(t)^2} |\gamma(\vb{k},t)|^2 -2 \frac{k^2}{a(t)^2} \, H(t)\, \gamma(\vb{k},t)^\ast \, \dot{\gamma}(\vb{k},t) 
- 2 \frac{k^2}{a(t)^2} \gamma(\vb{k},t)^\ast \, \ddot{\gamma}(\vb{k},t) + 2 \, H(t) \, \dot{\gamma}(\vb{k},t)^\ast \, \ddot{\gamma}(\vb{k},t) +|\ddot{\gamma}(\vb{k},t)|^2 \Bigg) \nonumber \\[0.3em]
&\!\!\overset{\mathrm{R.O.}}{=} \! \int\!\! \dd{t} \!\!\int\!\! \frac{ \dd[3]{\vb{k}} }{(2\pi)^3} \, a(t)^3 Z_{\mathrm{t}}^{(\alpha)}(t) \Bigg(\! \frac{1}{2} |\dot{\gamma} (\vb{k},t)|^2 - \frac{k^2}{a(t)^2}\frac{1}{2} |\gamma(\vb{k},t)|^2 \!  \Bigg) \! - \! \frac{\beta}{64 \pi G}\,  \! \int\!\! \dd{t} \!\!\int\!\! \frac{ \dd[3]{\vb{k}} }{(2\pi)^3} \, a(t)^3  \Bigg(\! 4 \frac{k^4}{a(t)^4} \, |\gamma(\vb{k},t)|^2 \nonumber \\
&\qquad + 4 H(t)^2 \, (1-\epsilon_{1Zt}^{(\alpha)}(t) + \tfrac{1}{4} \,\epsilon_{1Zt}^{(\alpha)}(t)^2 + 4 \tfrac{k^2}{a(t)^2}) \, |\dot{\gamma}(\vb{k},t)|^2 + 8 \frac{k^2}{a(t)^2} H(t) (1-\tfrac{1}{2}\, \epsilon_{1Zt}^{(\alpha)} (t) )\, \gamma(\vb{k},t)^\ast \, \dot{\gamma}(\vb{k},t) 
\Bigg) \nonumber \\[0.3em]
&\!\!\!\overset{\mathrm{I.B.P.}}{=}\! \int\!\! \dd{t} \!\!\int\!\! \frac{ \dd[3]{\vb{k}} }{(2\pi)^3} \, a(t)^3 Z_{\mathrm{t}}^{(\alpha)}(t) \Bigg( \frac{1}{2} |\dot{\gamma} (\vb{k},t)|^2 - \frac{k^2}{a(t)^2}\frac{1}{2} |\gamma(\vb{k},t)|^2   \Bigg)\!  -\!\frac{\beta}{64 \pi G}\, \! \! \int\!\! \dd{t} \!\!\int\!\! \frac{ \dd[3]{\vb{k}} }{(2\pi)^3}  a(t)^3  \Bigg(\! 4 H(t)^2 (1 \! - \! \tfrac{1}{2} \epsilon_{1Zt}^{(\alpha)}(t) )^2 \,|\dot{\gamma}(\vb{k},t)|^2 \nonumber \\
&\qquad + 4 \frac{k^2}{a(t)^2} \,\gamma(\vb{k},t)^{\ast}\, \Big[  \ddot{\gamma} (\vb{k},t) + H(t) (3-\epsilon_{1Zt}^{(\alpha)}(t))\, \dot{\gamma}(\vb{k},t) + \frac{k^2}{a(t)^2} \,\gamma(\vb{k},t)  \Big]\Bigg) \nonumber \\[0.3em]
&\! \!\overset{\mathrm{R.O.}}{=} \! \int\!\! \dd{t} \!\!\int\!\! \frac{ \dd[3]{\vb{k}} }{(2\pi)^3} \, a(t)^3 \Bigg( Z_{\mathrm{t}}^{(\alpha)}(t) \frac{1}{2} |\dot{\gamma} (\vb{k},t)|^2 - \frac{k^2}{a(t)^2}\frac{1}{2} |\gamma(\vb{k},t)|^2 - \frac{\beta}{64\pi G} 4 H(t)^2 (1-\tfrac{1}{2} \epsilon_{1Zt}^{(\alpha)}(t))^2\, |\dot{\gamma}(\vb{k},t)|^2  \Bigg) 
\nonumber \\[0.3em]
&=  \! \int\!\! \dd{t} \!\!\int\!\! \frac{ \dd[3]{\vb{k}} }{(2\pi)^3} \, a(t)^3 \, Z_{\mathrm{t}} (t) \Bigg( \frac{1}{2} \,|\dot{\gamma}(\vb{k},t)|^2 - c_{\mathrm{t}}(t)^2 \frac{k^2}{a(t)^2} \frac{1}{2} \, | \gamma(\vb{k},t) |^2\Bigg)\, .
\end{align}	
\end{widetext}
The quadratic action is now of the form \eqref{eq:Quadratic-Action-psi}, with effective kinetic amplitude and speed of sound given by
\begin{align}
	\label{eq:Z-t-Weyl-Starobinsky}
	Z_{\mathrm{t}} (t)&=  \frac{Z_{\mathrm{t}}^{(\alpha)}}{c_{\mathrm{t}}(t)^2} = \frac{\bar{\chi}(t) - 8 \, \beta\, H(t)^2 (1-\tfrac{1}{2} \epsilon_{1\chi}(t))^2 }{64\, \pi \, G}\, , \nonumber\\[.5em]
		\frac{1}{c_{\mathrm{t}}(t)^2} &= 1 - 8\,\beta\, \frac{H(t)^2 (1-\tfrac{1}{2}\epsilon_{1\chi}(t))^2}{\bar{\chi}(t)}\, .
\end{align}
As a consequence of the particular background geometry, we can expand all the previous quantities uniquely in terms of $\epsilon_{1H}(t)$; e.g., see \eqref{eq:H-to-eps}. Similarly, for tensor modes, the expanded kinetic amplitude and speed of sound up to N3LO are
\begin{align}
Z_{\mathrm{t}}(t) &= \frac{1}{96 G \pi \epsilon_{1H}(t)} \Big(
		1 - \frac{\beta}{3 \alpha} + \frac{5}{6} \epsilon_{1H}(t) 
		+ \frac{13 \beta}{18 \alpha} \, \epsilon_{1H}(t) \nonumber \\
		&\quad + \frac{2}{9} \epsilon_{1H}(t)^2 
		- \frac{125 \beta}{108 \alpha} \, \epsilon_{1H}(t)^2 
		- \frac{8}{27} \epsilon_{1H}(t)^3  \nonumber \\
		& \quad + \frac{937 \beta}{648 \alpha} \, \epsilon_{1H}(t)^3 
		+ \frac{2}{3} \epsilon_{1H}(t)^4 
		- \frac{653 \beta}{432 \alpha} \, \epsilon_{1H}(t)^4
		\Big), \nonumber \\[.5em]
c_{\mathrm{t}}(t) &= 1 + \frac{\beta}{6 \alpha} 
		- \frac{\beta}{2 \alpha} \, \epsilon_{1H}(t)
		+ \frac{23 \beta}{24 \alpha} \, \epsilon_{1H}(t)^2 
		- \frac{49 \beta}{36 \alpha} \, \epsilon_{1H}(t)^3  \nonumber \\
		&\quad + \frac{1225 \beta}{864 \alpha} \, \epsilon_{1H}(t)^4 \, ,
\end{align}
with the associated Hubble flow parameters
\begin{align}
\epsilon_{1Z}^{(\mathrm{t})}(t) &= 2 \epsilon_{1H}(t) - 2 \epsilon_{1H}(t)^2 
		- \tfrac{2 \beta}{\alpha} \, \epsilon_{1H}(t)^2 
		+ \tfrac{4}{3} \epsilon_{1H}(t)^3 \nonumber \\
		&\quad  + \tfrac{8 \beta}{\alpha} \, \epsilon_{1H}(t)^3, \nonumber \\[.5em]
\epsilon_{2Z}^{(\mathrm{t})}(t) &= -2 \epsilon_{1H}(t) + \tfrac{7}{3} \epsilon_{1H}(t)^2 
		+ \tfrac{2 \beta}{\alpha} \, \epsilon_{1H}(t)^2 
		- \tfrac{14}{9} \epsilon_{1H}(t)^3 \nonumber \\
		&\quad - \tfrac{37 \beta}{3 \alpha} \, \epsilon_{1H}(t)^3, \nonumber \\[.5em]
\epsilon_{3Z}^{(\mathrm{t})}(t) &= -2 \epsilon_{1H}(t) + \tfrac{8}{3} \epsilon_{1H}(t)^2 
		+ \tfrac{2 \beta}{\alpha} \, \epsilon_{1H}(t)^2 
		- \tfrac{4}{3} \epsilon_{1H}(t)^3 \nonumber \\
		&\quad - \tfrac{61 \beta}{3 \alpha} \, \epsilon_{1H}(t)^3, \nonumber \\[1em]
		\epsilon_{1c}^{(\mathrm{t})}(t) &= \tfrac{\beta}{\alpha} \, \epsilon_{1H}(t)^2 
		- \tfrac{4 \beta}{\alpha} \, \epsilon_{1H}(t)^3, \nonumber \\[.5em]
		\epsilon_{2c}^{(\mathrm{t})}(t) &= -4 \epsilon_{1H}(t) + \tfrac{26}{3} \epsilon_{1H}(t)^2 
		+ \tfrac{2 \beta}{\alpha} \, \epsilon_{1H}(t)^2 
		- \tfrac{61}{9} \epsilon_{1H}(t)^3 \nonumber \\
		&\quad - \tfrac{8 \beta}{\alpha} \, \epsilon_{1H}(t)^3, \nonumber \\[.5em]
		\epsilon_{3c}^{(\mathrm{t})}(t) &= -2 \epsilon_{1H}(t) + \tfrac{14}{3} \epsilon_{1H}(t)^2 
		+ \tfrac{\beta}{\alpha} \, \epsilon_{1H}(t)^2 
		+ \tfrac{4}{3} \epsilon_{1H}(t)^3 \nonumber \\
		&\quad - \tfrac{23 \beta}{6 \alpha} \, \epsilon_{1H}(t)^3.
\end{align}
Note that in the limit $\beta \to 0$, the functions for both modes reduce to the values of Starobinsky perturbations, i.e., \eqref{eq:Z-s-Starobinsky} and \eqref{eq:Z-t-Starobinsky}. All the corrections are expressed in terms of $H(t)$, $\bar{\chi}(t)$, and $\epsilon_{1\chi}(t)$, reflecting the purely geometric nature of this approach to inflation.

\section{Inflationary observables and comparison with observations}
\label{sec:Inflationary-Observations}

We summarize the formalism developed in \cite{Bianchi:2024qyp, Bianchi:2024jmn} for the computation of the primordial power spectrum  associated with any theory of inflation with quadratic action of the form \eqref{eq:Quadratic-Action-psi}. The primordial tensor and curvature perturbations that induce the seeds of the large-scale structure and the CMB temperature anisotropies are assumed to be quantum fields $\hat{\Psi} (\vb{x},t)$ initially in a Fock vacuum $\ket{0}$, defined by $\hat{a}(\vb{k}) \ket{0} = 0\,$, $\forall\, \vb{k}$, with bosonic creation and annihilation operators, $[\hat{a}(\vb{k}),\hat{a}^\dagger(\vb{k^\prime})] = (2\pi)^3 \,\delta^{(3)} (\vb{k}-\vb{k}^\prime)$. In Fourier space, the field is represented as a mode expansion 
\begin{equation}
   \hat{\psi} (\vb{k},t) = u(k,t) \,\hat{a}(\vb{k}) + u^\ast (k,t)\,\hat{a}^\dagger (-\vb{k})\, . 
\end{equation}
We consider a generalization of the Mukhanov-Sasaki variables, and perform a time reparametrization 
\begin{equation}
t\;\to \;y \equiv -k\, \tau = k\,    \hat{c}_\psi(t) \, \eta\, , 
\end{equation}
where $\tau=-\hat{c}_\psi(t) \, \eta$ generalizes the conformal time $\eta$ including the speed of sound of \eqref{eq:Quadratic-Action-psi}, and $\hat{c}_\psi(t)$ is defined in Appendix \ref{app:AppA}. Additionally, the mode functions are rescaled via,
\begin{equation}
u(k,t) \to \frac{y\, w(y)}{\sqrt{2\, k^3\, \mu(y)}}\, ,
\end{equation}
with $\mu(y) = (\hbar \, H(t)^2)^{-1}\, Z_\psi(t) \, c_\psi(t)\, \tilde{c}_\psi^2$, and $\tilde{c}_{\psi} = \hat{c}_\psi \, (a(t)\, H(t)\,\eta)$.  The canonical commutation relations become canonical Wronskian conditions for the mode functions, i.e., $w(y)\, w'{}^\ast (y) - w'(y)\,  w^\ast (y)  = -2 \,\ii$, while their equation of motion is given by
\begin{equation}
	\label{eq:EOM}
	w''(y)+\qty(1-\frac{2}{y^2} )w(y) = \frac{g(y)}{y^2}\,w(y)\,,
\end{equation}
where $g(y) = g_{1k} + g_{2k}\, \ln(y)\,+\,g_{3k}\, \ln(y)^2+\order{\epsilon^4}$ and the coefficients that depend on the background quantities $H(t)$, $Z_\psi(t)$, and $c_\psi(t)$, are given in \eqref{eq:g's-full-expression}. A derivation of \eqref{eq:EOM} is discussed in Appendix \ref{app:Log-Expansion}. The mode function $w_{\mathrm{qBD}}(y)$, associated with the quasi-Bunch-Davies vacuum $\ket{0}$, is the unique solution of the Wronskian condition and the mode equation \eqref{eq:EOM}, determined iteratively via the Green function method \cite{Stewart2001,Auclair2022,Bianchi:2024qyp,Ballardini:2024irx,Bianchi:2024jmn}. For this state, the late-time power spectrum at N3LO is given by
\begin{align}
	\label{eq:Power-BunchDavies}
	&\!\!\!\mathcal{P}_{\mathrm{qBD}}^{(\psi)} (k)\equiv\lim_{t\to \infty } \frac{k^3}{2\pi^2}  |u_{\mathrm{qBD}}(k,t)|^2  = \lim_{y\to 0^+ } \frac{\abs{y\,w_{\mathrm{qBD}}(y)}^2}{4\pi^2\,\mu(y)} \nonumber \\[.5em]
	&= \frac{\hbar\, H_\ast^2}{4\pi^2 Z^{(\psi)}_\ast c^{(\psi)3}_\ast} \Bigg[p_{0\ast} + p_{1\ast} \ln(\frac{k}{k_\ast} ) + p_{2\ast} \ln(\frac{k}{k_\ast})^2\nonumber \\
	&\hspace{2.5cm}+ p_{3\ast} \ln(\frac{k}{k_\ast})^3 \Bigg]\,,
\end{align}
where each of the quantities is evaluated at a pivot scale defined by $k_\ast \tau_\ast =-1$, following the logarithmic expansion described in Appendix \ref{app:Log-Expansion}, and the coefficients $p_{0\ast}$, $p_{1\ast}$, and $p_{2\ast}$, which depend on the Hubble flow parameters, are explicitly given in Tables. III, IV, V, and VI of \cite{Bianchi:2024qyp}. It is important to note that the generalized conformal time $\tau$ depends implicitly on the speed of sound $\hat{c}_{\psi}(t)$. As it is clear from \eqref{eq:Z-s-Weyl-Starobinsky} and \eqref{eq:Z-t-Weyl-Starobinsky}, scalar and tensor modes have different speeds of sound from each other. As a consequence, the same pivot scale is associated with two slightly different times for the freezing of scalar and of tensor modes. The precision prediction of the tensor-to-scalar ratio requires that we determine the relevant scale for scalars and for tensors consistently. As scalar modes have already been measured, we choose to work with the generalized time associated with scalar modes. To be more concrete, for tensor modes we denote the pivot scale defined by $k_\triangle \tau_\triangle = k_\triangle\, \hat{c}_{\mathrm{t}}(\eta_\triangle) \, \eta_\triangle =-1$, while for scalar modes we have $k_{\odot} \tau_{\odot} =k_\odot \, \hat{c}_{\mathrm{s}}(\eta_\odot) \, \eta_\odot =-1$, such that the fully expanded power spectra for both modes are given by
\begin{widetext}
\begin{align}
\mathcal{P}_{\mathrm{qBD}}^{(\mathrm{s})} (k) &=  \frac{\hbar\, H_\odot^2}{4\pi^2 Z^{(\mathrm{s})}_\odot c_{\mathrm{s}\,\odot}^3} \Bigg[p_{0\odot}^{(\mathrm{s})} + p_{1\odot}^{(\mathrm{s})} \ln(\frac{k}{k_\odot} ) + p_{2\odot}^{(\mathrm{s})} \ln(\frac{k}{k_\odot})^2 + p_{3\odot}^{(\mathrm{s})} \ln(\frac{k}{k_\odot})^3 \Bigg]\,, \label{eq:Power-Scalars-Full-Odot}\\[.5em] 
\mathcal{P}_{\mathrm{qBD}}^{(\mathrm{t})} (k) &=  \frac{\hbar\, H_\triangle^2}{4\pi^2 Z^{(\mathrm{t})}_\triangle c_{\mathrm{t}\,\triangle}^3} \Bigg[p_{0\triangle}^{(\mathrm{t})} + p_{1\triangle}^{(\mathrm{t})} \ln(\frac{k}{k_\triangle} ) + p_{2\triangle}^{(\mathrm{t})} \ln(\frac{k}{k_\triangle})^2 + p_{3\triangle}^{(\mathrm{t})} \ln(\frac{k}{k_\triangle})^3 \Bigg]\,. \label{eq:Power-Tensors-Full-triangle}
\end{align}
\end{widetext}

\begin{figure*}[ht]
	\centering
	\includegraphics[width=\linewidth]{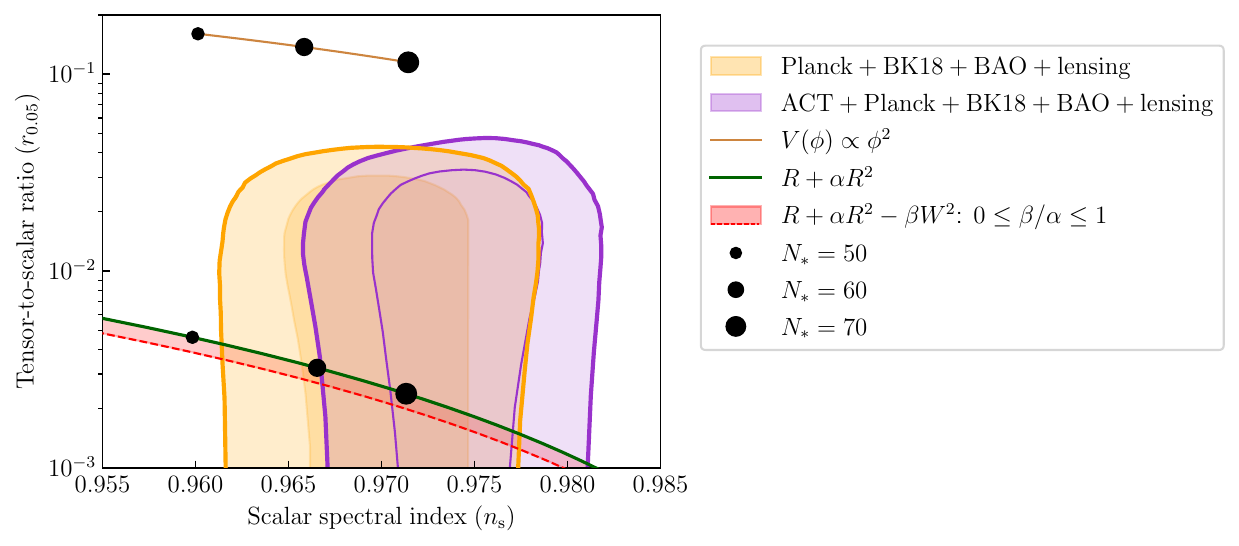}
	\caption{Marginalized joint 68\% and 95\% C.L. regions for $n_s$ and $r$ at $k = 0.05$ Mpc$^{-1}$ as reported by the Atacama Cosmology Telescope (ACT) Collaboration in their ACT Data Release 6 \cite{ACT:2025tim}, where constraints from Planck, BICEP2/Keck and CMB Lensing+Baryonic Acoustic Oscillations are also included. The results of Table~\ref{Tab:Results_Summaryv2} for Starobinsky inflation up to N3LO are shown in the solid green line, while the red shaded region indicates the possible values on the plane for a range $0\leq \beta/\alpha \leq 1$. For $\beta>0$, the predicted values of $r$ are lower than the ones for Starobinsky inflation. As argued in \cite{Zharov:2025evb}, larger values of $N_\ast$ can be taken into account in reheating constraints, allowing the model to remain consistent with the latest observational data.}
	\label{fig:rnsplane}
\end{figure*}
The explicit procedure to shift the expression for the power from one pivot to another has been outlined in Appendix B of \cite{Bianchi:2024qyp}. We include here a short summary: we implicitly assume that $\eta_\odot = \eta_\triangle$, which corresponds to relabel the evaluated coefficients, i.e., $H_\triangle$, $Z_\triangle$, $c_\triangle$, and $p_{n\,\triangle}$, in \eqref{eq:Power-Tensors-Full-triangle} and change $\triangle$ by $\odot$. Then, as $k_{\odot}/k_\triangle = \tau_\triangle/\tau_\odot = \hat{c}_{\mathrm{t}} (\eta_\odot) /\hat{c}_{\mathrm{s}}(\eta_\odot)$, one obtains
\begin{equation}
 \ln(\frac{k}{k_\triangle}) = \ln(\frac{k}{k_\odot}) +\ln(\frac{\hat{c}_{\mathrm{t}\,\odot}}{\hat{c}_{\mathrm{s}\,\odot}})\, .
\end{equation}
Consequently, the primordial power spectrum of tensor perturbations can be expressed as an expansion around the pivot scale $k_\odot$, given by
\begin{widetext}
\begin{equation}
	\label{eq:Power-Tensors-Full-Odot}
\mathcal{P}_{\mathrm{qBD}}^{(\mathrm{t})} (k) =  \frac{\hbar\, H_\odot^2}{4\pi^2 Z^{(\mathrm{t})}_\odot c_{\mathrm{t}\,\odot}^3} \Bigg[p_{0\odot}^{(\mathrm{t})} + p_{1\odot}^{(\mathrm{t})} \qty[\ln(\frac{k}{k_\odot} ) + \ln(\frac{\hat{c}_{\mathrm{t}\,\odot}}{\hat{c}_{\mathrm{s}\,\odot}})] + p_{2\odot}^{(\mathrm{t})}\qty[ \ln(\frac{k}{k_\odot})+ \ln(\frac{\hat{c}_{\mathrm{t}\,\odot}}{\hat{c}_{\mathrm{s}\,\odot}})]^2 + p_{3\odot}^{(\mathrm{t})} \qty[ \ln(\frac{k}{k_\odot}) \ln(\frac{\hat{c}_{\mathrm{t}\,\odot}}{\hat{c}_{\mathrm{s}\,\odot}})]^3 \Bigg].
\end{equation}
\end{widetext}
Note that the ratio between the speeds of sound for tensor and scalar modes satisfies the relation
\begin{equation}
\frac{c_{\mathrm{t}\odot} }{c_{\mathrm{s}\odot}} = 1 + \frac{\beta}{6\, \alpha} - \frac{\beta}{2\, \alpha}  \epsilon_{1H\odot} + \frac{19}{24} \frac{\beta}{\alpha}  \epsilon_{1H\odot}^2 - \frac{31}{36}  \frac{\beta}{\alpha} \epsilon_{1H\odot}^3 + \order{\epsilon^4}\, .
\end{equation}
To facilitate the comparison with observational constraints, note that the logarithm of the fully expanded primordial power spectrum \eqref{eq:Power-BunchDavies} can be cast in the following form: 
\begin{align}
	\label{eq:Log-Power}
	&\ln(\mathcal{P}^{(\mathrm{s,t})}_{\mathrm{qBd}}(k)) = \ln(\mathcal{A}_{\mathrm{s,t}}) + \theta_{\mathrm{s,t}} \ln(\frac{k}{k_\odot}) \nonumber \\
	&\quad + \frac{\mathfrak{a}_{\mathrm{s,t}}}{2!} \ln(\frac{k}{k_\odot})^2 + \frac{\mathfrak{b}_{\mathrm{s,t}}}{3!} \ln(\frac{k}{k_\odot})^3 + \order{\textrm{N4LO}} \, ,
\end{align}
where the identifications with the coefficients $p_{n\,\odot}$ are valid, in terms of Hubble flow parameters, up to $\order{\epsilon^3}$:
\begin{subequations}
	\label{eq:Power-Law-p'ns}
	\begin{align}
		\mathcal{A}_{\mathrm{s,t}} &=  \frac{\hbar\, H_\odot^2}{4\pi^2 Z^{(\mathrm{s,t})}_\odot c_{\mathrm{s,t}\,\odot}^3} \; p_{0\odot}^{(\mathrm{s,t})} \, , \\
		\theta_{\mathrm{s,t}} &= \frac{p_{1\odot}^{(\mathrm{s,t})} }{   p_{0\odot}^{(\mathrm{s,t})} } \, ,\\
		\mathfrak{a}_{\mathrm{s,t}} &= - \frac{ p_{1\odot}^{(\mathrm{s,t})2}  }{p_{0\odot}^{(\mathrm{s,t})2}} \qty(  1 - 2 \frac{ p_{0\odot}^{(\mathrm{s,t})} \, p_{2\odot}^{(\mathrm{s,t})}   }{  p_{1\odot}^{(\mathrm{s,t})2} }   ) \, , \\
		\mathfrak{b}_{\mathrm{s,t}} &= \qty(2 \, \frac{p_{1\odot}^{(\mathrm{s,t})3}}{ p_{0\odot}^{(\mathrm{s,t})3}  } ) \qty(  1 -  3 \, \frac{p_{0\odot}^{(\mathrm{s,t})}\, p_{2\odot}^{(\mathrm{s,t})} }{ p_{1\odot}^{(\mathrm{s,t})2}  }  + 3\, \frac{p_{0\odot}^{(\mathrm{s,t})2}\, p_{3\odot}^{(\mathrm{s,t})} }{p_{1\odot}^{(\mathrm{s,t})3}}  )\,. 
	\end{align}
\end{subequations}

The above quantities describe deviations from an exact power law, which is often employed as a phenomenological template for observational constraints. These beyond-power-law quantities are the amplitude $\mathcal{A}$ at the pivot mode $k_\odot$, together with its logarithmic derivatives: the spectral tilt $\theta$, the running of the tilt $\mathfrak{a}$, and the running-of-the-running of the tilt $\mathfrak{b}$, which are formally  defined as
\begin{subequations}
	\label{eq:Power-Law-definition}
	\begin{align}
		\mathcal{A}_{\mathrm{s,t}} &\equiv \mathcal{P}_{\mathrm{qBD}}^{({\mathrm{s,t}})}(k_\odot)\, , \\[.5em]
		\theta_{\mathrm{s,t}} &\equiv  \eval{ \dv{ \ln( \mathcal{P}_{\mathrm{qBD}}^{(\mathrm{s,t})} (k) )}{\ln(k)}  }_{k=k_\odot} \, , \\
		\mathfrak{a}_{\mathrm{s,t}}  &\equiv \eval{ \dv[2]{ \ln( \mathcal{P}_{\mathrm{qBD}}^{(\mathrm{s,t})} (k) )}{\ln(k)}  }_{k=k_\odot}\, , \\
		\mathfrak{b}_{\mathrm{s,t}}  &\equiv \eval{ \dv[3]{ \ln( \mathcal{P}_{\mathrm{qBD}}^{(\mathrm{s,t})} (k) )}{\ln(k)}  }_{k=k_\odot}\, .
	\end{align}
\end{subequations}
Consequently, we can derive explicit expressions for the phenomenological quantities \eqref{eq:Power-Law-definition} directly by evaluating \eqref{eq:Power-Law-p'ns} and use them to compare with observational constraints. The explicit expressions of these quantities, expressed in terms of $\epsilon_{1H\odot}$, can be found in \cite{ZenodoR2W2}. Furthermore, one special feature of $R+\alpha\, R^2-\beta\, W^2$ inflation is the fact that the quantities in \eqref{eq:Power-Law-p'ns} can be expressed in terms of the scalar spectral tilt $\theta_\mathrm{s}\equiv n_\mathrm{s}-1$, one of the most precisely measured observables in inflationary cosmology. To see this, note that evaluating the expansions \eqref{eq:Epsilon_Expansion_234}, we find
that the scalar tilt at $k_\odot$ takes the form
\begin{align}
\label{eq:ns-in-terms-epsilon}
n_{\mathrm{s}}-1 &=
-4\,\epsilon_{1H\odot}
- \frac{28}{3}\,\epsilon_{1H\odot}^2
- 8\, C\,\epsilon_{1H\odot}^2 - \frac{880}{9}\,\epsilon_{1H\odot}^3\nonumber \\
&\quad - 44\, C\,\epsilon_{1H\odot}^3
- 16\, C^2\,\epsilon_{1H\odot}^3
+ \frac{28\pi^2}{3}\,\epsilon_{1H\odot}^3\nonumber \\
&\quad 
+ \frac{2\,\beta}{3\,\alpha}\,\epsilon_{1H\odot}^3 + \order{\mathrm{N4LO}}\, ,
\end{align}
where $C = \gamma_E + \ln(2) - 2 \simeq -0.730$. Hence, as any quantity can be written in terms of $\epsilon_{1H\odot}$ using \eqref{eq:Epsilon_Expansion_234}, the above expression can be used to rewrite each of them as a series in $n_{\mathrm{s}}-1$. A direct advantage of doing so comes from the fact that $n_\mathrm{s}-1$ is frame independent, which allows us to write the inflationary predictions as a set of specific relations between different observables as measured at a pivot $k_\odot$. In Table~\ref{Tab:Results_Summaryv2}, we report the explicit expressions for the quantities in \eqref{eq:Power-Law-p'ns} expanded in powers of $n_{\mathrm{s}}-1$. The resulting relation between $r$ and $n_{\mathrm{s}}$ is confronted with observational constraints in Fig.~\ref{fig:rnsplane}.

\begin{table}
	\caption{Relations between inflationary observables in $R+\alpha\, R^2 - \beta\, W^2$ inflation. The observables are defined in \eqref{eq:Power-Law-definition}, with $r = \mathcal{A}_{\mathrm{t}}/\mathcal{A}_{\mathrm{s}}$, $n_{\mathrm{s}} = \theta_{\mathrm{s}}+1$, $n_{\mathrm{t}} = \theta_{\mathrm{t}}$. The amplitudes $\mathcal{A}_{\mathrm{s}}$ and $\mathcal{A}_{\mathrm{t}}$ are reported up to NLO, while all other observables are reported up to N3LO.}
	\label{Tab:Results_Summaryv2}
	\begin{ruledtabular}
		\begin{tabular}{c l}
			\rule{0pt}{3ex} Observable  & Predictions from $R+\alpha\, R^2 - \beta\, W^2$  \\[.5em] 
			\hline
			\rule{0pt}{4ex}$\mathcal{A}_{\mathrm{s}}$ & $\dfrac{2\, G\, \hbar  }{9\, \pi\,  \alpha } \dfrac{1}{(n_\mathrm{s}-1)^2} \Big(1-\frac{5}{12} \left(n_s-1\right) \Big) $ \\[.7em] 
			\hline
			\rule{0pt}{4ex}$\mathcal{A}_{\mathrm{t}}$ & $\dfrac{2\, G\, \hbar  }{3\,\pi \, \alpha }  \Big(1 - \tfrac{1}{6}\tfrac{\beta}{\alpha }+\frac{3}{4} \big(1 -\tfrac{1}{3}\tfrac{\beta}{\alpha} \big)\left(n_s-1\right) \Big) $\\[.7em] 
			\hline
			\rule{0pt}{3ex}$r$& $3 \, (1-\tfrac{1}{6}\tfrac{\beta}{\alpha})  (n_{\mathrm{s}} - 1)^2 + \tfrac{7}{2} (1-\tfrac{23}{84}\tfrac{\beta}{\alpha} ) (n_{\mathrm{s}} - 1)^3$ \\[.5em] 
			\hline 
			\rule{0pt}{3ex}$n_{\mathrm{s}}$ & $n_\mathrm{s}$ \\[.5em] 
			\hline
			\rule{0pt}{3ex}$n_{\mathrm{t}} $ & $-\tfrac{3}{8} (1-\tfrac{1}{6}\tfrac{\beta}{\,\alpha}) (n_{\mathrm{s}} - 1)^2  -\tfrac{1}{16} (1-\tfrac{11}{12}\tfrac{\beta}{\alpha})  (n_{\mathrm{s}} - 1)^3$  \\[.5em] 
			\hline
			\rule{0pt}{3ex}$r+8\, n_{\mathrm{t}}$ & $3\, (1-\tfrac{1}{6}\tfrac{\beta}{\alpha}) (n_{\mathrm{s}} - 1)^3 $
			\\[.5em]	
			\hline
			\rule{0pt}{3ex}$\mathfrak{a}_{\mathrm{s}}$ & $-\tfrac{1}{2}  (n_{\mathrm{s}} - 1)^2 + \tfrac{5}{48}  (n_{\mathrm{s}} - 1)^3$  \\[.5em] 
			\hline
			\rule{0pt}{3ex}$\mathfrak{a}_{\mathrm{t}}$ &			 $-\tfrac{3}{8}(1-\tfrac{1}{6}\tfrac{\beta}{\alpha})  (n_{\mathrm{s}} - 1)^3$  \\[.5em] 			 
			\hline
			\rule{0pt}{3ex}$\mathfrak{b}_{\mathrm{s}}$ & $\tfrac{1}{2}  (n_{\mathrm{s}} - 1)^3$ \\[.5em] 
			\hline
			\rule{0pt}{3ex}$\mathfrak{b}_{\mathrm{t}}$ & $0$ 
		\end{tabular}
	\end{ruledtabular}
\end{table}

It is important to remark that the ratio $\beta/\alpha$ only enters as a N3LO correction in \eqref{eq:ns-in-terms-epsilon}, and that the main difference between the pivot defined as $k_\odot \approx a_\odot H_\odot/c_{\odot}^{(\mathrm{s})}$ compared to the usual pivot $k_\ast \approx a_\ast H_\ast$ comes from the speed of sound for scalar perturbations, which is already N2LO suppressed, i.e., $c_{\odot}^{(\mathrm{s})} = 1 + (\beta/6\alpha)\, \epsilon_{1H\odot}^2 + \order{\epsilon^3}$. As a consequence, to report other quantities that do not need to be predicted with such high level of precision (such as the number of e-foldings $N_\ast$), and to facilitate the comparison with other results in the literature, we provide also NLO expressions that do not depend on the contributions of the speed of sound and therefore use the pivot $k_\ast$. Defining as usual $t_\ast$ as the time when the mode $k_\ast$ freezes, i.e., $a(t_\ast) H(t_\ast)\equiv k_\ast$, and $t_{\mathrm{end}}$ as the time where inflation ends, i.e., $\epsilon_{1H}(t_{\mathrm{end}})\equiv 1$, and using  \eqref{eq:Ns-via-epsilon}, the number of e-foldings $N_\ast$ can be expressed as
\begin{align}
	\label{eq:Ns-via-ns}
	N_{\ast} &= \int_{t_\ast}^{t_{\textrm{end}}} H(t) \dd{t} = - \int_{\epsilon_{1H\ast}}^{1} \frac{\dd{\epsilon_{1H}}}{ \epsilon_{1H}\;\, \epsilon_{2H}\big(\epsilon_{1H}\big)  }  \nonumber \\[.5em]
    &=\frac{2}{1-n_{\mathrm{s}}} + \frac{1}{12} \ln(\frac{1}{1-n_{\mathrm{s}}}) - d_0 + \order{1-n_\mathrm{s}} \, ,
\end{align}
where $d_0 \approx 0.02$. A detailed derivation of the above formula is given in Appendix~\ref{app:Computation_N_odot}. Similarly, the dimensionless gap defined in \eqref{eq:delta-definition}, which plays a key role in postinflationary reheating, is given by
\begin{align}
\label{eq:delta_via_ns}
\Delta_\ast &=N_\ast - \ln(H_\ast/H_{\mathrm{end}}) \nonumber \\
&=\frac{2}{1-n_\mathrm{s}} - \frac{5}{12} \, \ln(\frac{1}{1-n_{\mathrm{s}}}) - q_0 +\order{1-n_{\mathrm{s}}}\, ,
\end{align}
where $q_0 = d_0+\ln(28848/13583)\approx 0.77$, and where we have used the Padé approximant method to derive $H_{\mathrm{end}} \approx \frac{13583}{86544} \, \frac{1}{\sqrt{\alpha}}$, as shown in \eqref{eq:H_end}.

The set of relations reported in Table~\ref{Tab:Results_Summaryv2}, and the expressions for $N_\ast$ and $\Delta_\ast$ in \eqref{eq:Ns-via-ns} and \eqref{eq:delta_via_ns}  allow us to compute precision predictions of $R+\alpha\, R^2-\beta\, W^2$ inflation by only fixing or measuring the value of $n_{\mathrm{s}}$ at one given pivot scale. We illustrate this statement in Table~\ref{Tab:Numerical-Results} where, assuming fiducial values for the observables $n_{\mathrm{s}}$ and $\mathcal{A}_{\mathrm{s}}$, we compute all the other primordial observables with N3LO precision. 

\begin{table}[h]
	\caption{Predicted values of observables in $R+\alpha\, R^2-\beta \, W^2$ inflation, assuming a fiducial value for $n_\mathrm{s}$ and $\mathcal{A}_{\mathrm{s}}$. }
	\label{Tab:Numerical-Results}
	\begin{ruledtabular}
		\begin{tabular}{c l}
			\rule{0pt}{3ex}Quantity & Value for fiducial  $n_{\mathrm{s}} = 0.97$,  $\mathcal{A}_{\textrm{s}} =2.1 \times 10^{-9}$ \\[.5em] 
			\hline
			\rule{0pt}{4ex}$\alpha $ & $  \dfrac{2\, G \, \hbar}{ 9\, \pi\,  \mathcal{A}_{\mathrm{s}}  \, (n_\mathrm{s}-1)^2} \approx 3.7 \times 10^{10} \, G \hbar$  \\[.7em] 
            \hline
            \rule{0pt}{4ex}$N_{\ast} $ & $ 66.9$  \\[.7em] 
            \hline
            \rule{0pt}{4ex}$\Delta_{\ast} $ & $ 64.4$  \\[.7em] 
            \hline
			\rule{0pt}{3ex}$\mathcal{A}_{\mathrm{t}} $ & $ 7.7\times 10^{-12}\, \times \Big(1 - 0.163\, \tfrac{\beta}{\alpha }\Big) $\\[.5em] 
			\hline
			\rule{0pt}{3ex}$r $ & $2.6\times 10^{-3}\,\times \Big(1 -0.163 \, \frac{\beta}{\alpha} \Big) $ \\[.5em] 
			\hline 
			\rule{0pt}{3ex}$n_{\mathrm{t}}  $ & $ -3.4 \times 10^{-4} \,\times \Big( 1  - 0.163\, \frac{\beta}{\alpha}\Big) $  \\[.5em] 
			\hline
			\rule{0pt}{3ex}$r+8\, n_{\mathrm{t}} $ & $ -8.1 \times 10^{-5} \,\times (1-0.167\, \tfrac{\beta}{\alpha})$
			\\[.5em]	
			\hline
			\rule{0pt}{3ex}$\mathfrak{a}_{\mathrm{s}} $ & $- 4.5 \times 10^{-4}$  \\[.5em] 
			\hline
			\rule{0pt}{3ex}$\mathfrak{a}_{\mathrm{t}}$ & $-1.0 \times  10^{-5} \,\times \Big(1-0.167\, \frac{\beta}{\alpha }\Big)$  \\[.5em] 			 
			\hline
			\rule{0pt}{3ex}$\mathfrak{b}_{\mathrm{s}}$ & $-1.4 \times 10^{-5}$ 
		\end{tabular}
	\end{ruledtabular}
\end{table}

\section{Discussion}
\label{sec:Discussion}

We studied the inflationary predictions of the most general diffeomorphism-invariant effective theory of geometry, up to quadratic curvature terms, with an effective Lagrangian density $-2\Lambda + R + \alpha\, R^2 - \beta\, W^2$. The scalaron arising from the $R^2$ term is stable and is treated as a genuine propagating degree of freedom; its quantization, together with that of the tensor modes, follows the standard procedure. In contrast, a standard treatment of the $\beta\,W^2$ term would result in the appearance of spurious higher-derivative tensor ghosts. In our framework, we capture the effective corrections to the physical modes due to the $\beta\,W^2$ term by adopting a reduction of order method \cite{Bhabha:1946zz,Simon:1990ic}, analogous to the Landau-Lifshitz formulation of the Lorentz-Abraham-Dirac radiation-reaction force \cite{Dirac:1938,Landau:1975}. This reduction of order, consistent with the logic of effective field theory, yields  primordial initial conditions with $\beta\,W^2$ corrections to the quasi-Bunch-Davies vacuum that are analytic in the coupling $\beta$ by construction, providing self-consistent predictions that continuously connect to the $R+\alpha R^2$ sector as $\beta \to 0$. In this context, we characterized primordial perturbations in $R+\alpha\, R^2-\beta\, W^2$ inflation, providing precision N3LO predictions obtained via Green's function method, together with the reduction of order method. These results extend the previous precision results obtained in \cite{Bianchi:2024qyp} for Starobinsky inflation, $R+\alpha\, R^2$, which provides a de facto one-parameter reference for evaluating the evidence of other inflationary models \cite{Martin:2013tda, Kallosh:2025ijd,Martin:2024qnn,Martin:2010hh,Ballardini:2024ado,Sorensen:2024ezb}.

%The reduction-of-order method applied to the Weyl-squared term $\beta\,W^2$ allows us to provide a self-consistent analysis of the most general diffeomorphism-invariant effective theory of geometry, up to quadratic curvature terms, with an effective Lagrangian of the form $-2\Lambda + R + \alpha\, R^2 - \beta\, W^2$. 

As in any effective theory, the coupling constants have to be measured before predictions can be extracted from the theory. Cosmological observations of the amplitude of scalar perturbations $\mathcal{A}_{\mathrm{s}}$ and of the scalar tilt $n_{\mathrm{s}}$ provide a measurement of the coupling constant $\alpha$,
\begin{equation}
\frac{\alpha}{G\, \hbar} \approx \frac{2}{9\, \pi} \frac{1}{ (n_{\mathrm{s}}-1)^2\mathcal{A}_{\mathrm{s}} }  \approx  3.7 \,\times\, 10^{10} \, ,
\end{equation}
which is compatible with the assumed hierarchy of scales $G\hbar\ll \alpha\ll \Lambda^{-1}$. Furthermore, a  detection of primordial gravitational waves and, specifically, a sufficiently precise measurement of the tensor-to-scalar ratio $r$, would provide a measurement of the ratio $\beta/\alpha$ as
\begin{equation}
\label{eq:beta-alpha}
\frac{\beta}{\alpha} = 6\left(1 -\frac{r}{3(n_{\mathrm{s}}-1)^2}\right)  + \order{\mathrm{N3LO}}\, , 
\end{equation}
which is treated at linear order  $\order{\beta/\alpha}$ in this paper, assuming $G\hbar\ll \beta\ll \alpha$ and adopting a self-consistent reduction of order of higher-derivative terms. We note that the value \eqref{eq:beta-alpha} of the $\beta/\alpha$ correction to the tensor-to-scalar ratio $r$ reproduces the leading order results computed in \cite{Deruelle:2012xv,Myung:2015vya,Salvio:2017xul,Anselmi2020,Kubo:2025jla} using other approaches, confirming its robustness. Nonetheless, using Green's function method for effective theories of inflation developed in \cite{Bianchi:2024qyp, Bianchi:2024jmn}, we computed the inflationary observables that characterize the primordial power spectrum of scalar and tensor modes up to N3LO in the Hubble-flow expansion, as reported in Table~\ref{Tab:Results_Summaryv2}. These expressions yield precise predictions extending previous results beyond leading order, such as deviations from the slow-roll consistency condition and explicit values for the running of the spectral tilts, as illustrated in Table~\ref{Tab:Numerical-Results}, which can be tested with next-generation CMB and gravitational wave observatories.

%It is also interesting to remark that, at least up to $\order{\mathrm{N3LO}}$, a simple relation between the dimensionless ratio $\beta/\alpha$ and higher order inflationary observables holds, namely
%\begin{equation}
%\frac{\beta}{\alpha} = 6 - \frac{r+8\, n_\mathrm{t}}{\mathfrak{b}_{\mathrm{s}}}  + \order{\mathrm{N4LO}}\, . 
%\end{equation}

Most of the currently available analyses of Starobinsky inflation are done in the Einstein frame, i.e., using an inflaton field potential $V(\varphi^{(E)})$ and a number of e-foldings $N_\ast^{(E)}$ defined with respect to the Einstein-frame auxiliary metric $g^{(E)}_{\mu\nu}$ obtained via the field redefinition $g_{\mu\nu} \to (g^{(E)}_{\mu\nu},\varphi^{(E)})$. In this paper we worked consistently in a geometric framework (Jordan frame) where there is no additional inflaton field. Specifically, the two quantities $N_\ast$ and $\Delta_\ast$ are defined directly with respect to the physical metric $g_{\mu\nu}$, assuming minimal coupling of Standard Model matter fields. While each of the results on inflationary observables reported in Table~\ref{Tab:Results_Summaryv2} are frame independent as they are expressed in a relational form in terms of $n_{\mathrm{s}}$, the values of $N_\ast$ and $\Delta_\ast$ explicitly depend on the choice of frame. For instance, the difference between the number of e-foldings in the Jordan and the Einstein frames is given by \cite{Racioppi:2021jai}
\begin{equation}
\label{eq:Jordan-Einstein}
N_\ast-N_\ast^{(E)}=\dfrac{1}{2} \ln(\frac{F_{\textrm{end}}}{F_\ast})\, ,
\end{equation}
where $F$ is the conformal factor relating the two metrics, i.e., $g_{\mu\nu}^{(E)} = F (\varphi^{(E)}) \, g_{\mu\nu}$. Typically, for $N_\ast = 55$, one has $N_\ast^{(E)} \approx 53.2$, which is roughly 3\% smaller.

It would be interesting to extend the analysis of $R+\alpha\, R^2-\beta\, W^2$ inflation presented in this paper to take into account observational constraints from the postinflationary reheating epoch. In fact, given the recent results from ACT DR6 \cite{ACT:2025tim} and other planned cosmological observatories \cite{Euclid:2023shr,LiteBIRDCollaboration2023}, a detailed characterization of the reheating epoch becomes increasingly relevant. We stress that the description of the reheating phase is particularly sensitive to the choice of frame \cite{Appleby:2009uf,Kannike:2015apa,vandeBruck:2016leo,Nandi:2019xlj,Nandi:2019xve,Dorsch:2024nan,Ketov:2025nkr}. Specifically, the rescaled reheating parameter $R_{\mathrm{reh}}$, which is used to characterize the reheating epoch \cite{Martin:2006rs,Martin:2010kz,Martin:2010hh,Martin:2014nya,Martin:2024qnn}, is related to the inflationary observables via the matching condition $N_\ast = N_0 + \ln(R_{\mathrm{reh}}) + \ln(H_\ast/H_{\mathrm{end}})$ \cite{Zharov:2025evb}, where $N_0\approx 61.1$ describes the number of e-foldings from the end of reheating until today, assuming typical values of the Standard Model thermal history. This matching condition can be written in terms of the parameter $\Delta_\ast$ computed in \eqref{eq:delta_via_ns}. We find that the rescaled reheating parameter $R_{\mathrm{reh}}$ is therefore constrained by current observations thanks to the relation
\begin{align}
\ln(R_{\mathrm{reh}})  &=\Delta_\ast - N_0\, .
\end{align}
This relation is a distinctive prediction of $R+\alpha\, R^2 - \beta\, W^2$ inflation in the geometric frame, as $\Delta_{\ast}$ is completely fixed by the gravitational dynamics and can be expressed in terms of $n_{\mathrm{s}}-1$, as shown in \eqref{eq:delta_via_ns}. For a fiducial value $n_{\mathrm{s}}=0.97$ for the scalar tilt, we have $\Delta_\ast\approx 64.4$ and therefore the prediction  $\ln(R_{\mathrm{reh}})\approx 3.3$. This estimate is consistent with a recent analysis of reheating in Starobinsky inflation using the Einstein frame \cite{Zharov:2025evb}. Other extensions of Starobinsky inflation have been proposed recently in light of the current ACT DR6 results \cite{Drees:2025ngb,Yin:2025rrs,Yogesh:2025wak,Addazi:2025qra,Choudhury:2025vso,Wolf:2025ecy,Frolovsky:2025iao}, and it would be interesting to compare the constraints that arise in these different approaches.

Another important direction for future work is the extension of the framework presented in this paper to include a preinflationary epoch. In fact, the transition from a quantum geometry phase to the effective field theory phase can leave observable imprints on the largest scales of the CMB, departing from the quasi–Bunch-Davies vacuum and resulting in features in the primordial power spectrum  \cite{Chluba:2015bqa,Bianchi:2024jmn,Euclid:2023shr}. Encoding the effects of the preinflationary epoch into squeezed vacua for the effective geometric theory of $R+\alpha\, R^2-\beta\, W^2$ inflation opens up a promising window into the phenomenology of quantum gravity. In fact, in the context of nonperturbative approaches to quantum geometry \cite{Ashtekar2021,Ashtekar:2011ni,Bianchi:2010zs,Gozzini:2019nbo,Gielen:2013kla,Dittrich:2021gww,Agullo:2023rqq}, the effects of the Weyl-squared term \cite{Borissova:2022clg,Dittrich:2023rcr}, and of other possible quantum geometry effects in cosmology \cite{Gielen:2025jcb,Oriti:2025lwx,MenaMarugan:2024vyy,MenaMarugan:2024zcv}, are currently being considered. The precision analysis of the effective theory of quantum geometry, presented here for the inflationary phase, can lead to the falsification of predictions from quantum gravity and quantum geometry, allowing next-generation CMB and gravitational-wave experiments to provide stringent tests of quantum gravitational physics.

\begin{acknowledgments}
M.G. is supported by the \href{https://anid.cl}{Agencia Nacional de Investigación y Desarrollo} (ANID) and \href{http://www.fulbright.cl/}{ Fulbright Chile} through the Fulbright Foreign Student Program and ANID BECAS/Doctorado BIO Fulbright-ANID 56190016.~E.B. acknowledges support from the National Science Foundation, Grants No. PHY-2207851 and No. PHY-2513194. This work was made possible through the support of the ID 62312 grant from the John Templeton Foundation, as part of the project \href{https://www.templeton.org/grant/the-quantum-information-structure-of-spacetime-qiss-second-phase}{``The Quantum Information Structure of Spacetime'' (QISS)}. The opinions expressed in this work are those of the authors and do not necessarily reflect the views of the John Templeton Foundation.
\end{acknowledgments}

% \bibliography{bib-inflation}
%\newpage
%\clearpage

\appendix

% %
% \begin{center}
%     \textbf{-- End Matter --}
% \end{center}
% %
\onecolumngrid 
%\begin{widetext}
\section{Generalized conformal time}
\label{app:AppA}
As shown in \cite{Bianchi:2024qyp}, the standard conformal time $\eta$ in a quasi-de Sitter background up to N3LO is given by
\begin{align}
	\eta^{(3)} &= -\frac{1}{a(t) H(t)} \bigg( 1 + \epsilon_{1H}(t) + \epsilon_{1H}^2(t) 
	- \epsilon_{1H}(t) \epsilon_{2H}(t) - 3 \epsilon_{1H}^2(t) \epsilon_{2H}(t) 
	+ \epsilon_{1H}(t) \epsilon_{2H}^2(t) + \epsilon_{1H}(t) \epsilon_{2H}(t) \epsilon_{3H}(t)+ \epsilon_{1H}^3(t)  \bigg). \label{eq:eta-qdS-full}
\end{align}
The generalized conformal time $\tau$ is given by,
\begin{align}
	\tau^{(3)}&\equiv \frac{\tilde{c}_\psi(t)}{a(t)H(t)} \nonumber\\
	&=  -\frac{ c_\psi(t)}{a(t)H(t)} \Big[  1 + \epsilon_{1H}(t)  - \epsilon_{1c}(t)  + \epsilon_{1H}(t)^2 - \epsilon_{1H}(t) \epsilon_{2H}(t)  - 2 \epsilon_{1c}(t) \epsilon_{1H}(t)  + \epsilon_{1c}(t)\epsilon_{2c}(t) + \epsilon_{1c}(t)^2   \nonumber \\
	&\quad 
	+\epsilon_{1H}(t)^3 + \epsilon_{1H}(t) \epsilon_{2H}(t) \epsilon_{3H}(t)  - 
	3 \epsilon_{1H}(t)^2 \epsilon_{2H}(t) + \epsilon_{1H}(t) \epsilon_{2H}(t)^2 - \epsilon_{1c}(t) \epsilon_{2c}(t)^2 + 
	3 \epsilon_{1c}(t) \epsilon_{1H}(t) \epsilon_{2H}(t)  \nonumber \\
	&\quad - 
	3 \epsilon_{1c}(t)^2 \epsilon_{2c}(t)  +  
	3 \epsilon_{1c}(t) \epsilon_{1H}(t) \epsilon_{2c}(t)  - 
	\epsilon_{1c}(t) \epsilon_{2c}(t) \epsilon_{3c}(t) 	-\epsilon_{1c}(t)^3 + 3 \epsilon_{1c}(t)^2 \epsilon_{1H}(t) - 
	3 \epsilon_{1c}(t) \epsilon_{1H}(t)^2  \Big]\,. \label{eq:tau-full}
\end{align}
It can be expressed as $\tau(t) =\hat{c}_\psi (t) \, \eta(t)$, where
\begin{align}
	\hat{c}_\psi (t) &\equiv c_\psi(t)\, \Big\{ 1 - \epsilon_{1c}(t) + \epsilon_{1c}(t)^2 - \epsilon_{1c}(t)^3 
	- \epsilon_{1c}(t) \epsilon_{1H}(t) 
	+ 2 \epsilon_{1c}(t)^2 \epsilon_{1H}(t) 
	- \epsilon_{1c}(t) \epsilon_{1H}(t)^2 
	+ \epsilon_{1c}(t) \epsilon_{2c}(t) \nonumber\\ 
	&\quad - 3 \epsilon_{1c}(t)^2 \epsilon_{2c}(t) 
	+ 2 \epsilon_{1c}(t) \epsilon_{1H}(t) \epsilon_{2c}(t) 
	- \epsilon_{1c}(t) \epsilon_{2c}(t)^2 
	+ 2 \epsilon_{1c}(t) \epsilon_{1H}(t) \epsilon_{2H}(t) 
	- \epsilon_{1c}(t) \epsilon_{2c}(t) \epsilon_{3c}(t)
	\Big\}. \label{eq:c-hat-def}
\end{align}

\section{Logarithmic expansion and fully expanded power spectrum around a pivot}
\label{app:Log-Expansion}
We summarize some of the steps in the derivation of the mode equation for $w(y)$, following \cite{Bianchi:2024qyp,Bianchi:2024jmn}. First, note that for the original mode $u(k,t)$ satisfies the mode equation
\begin{equation}
	\label{eq:EoM-u(t)}
	\ddot{u}(k,t) + (3-\epsilon_{Z1}(t)) H(t) \dot{u}(k,t) + c_\psi(t)^2\frac{k^2}{a(t)^2}  \, u(k,t) = 0\,.
\end{equation} 
Under the map $u(k,t)\to w(y)$, the above equation reads
\begin{equation}
	\label{eq:EoM-v(x)-aH}
	w''(y) +  \left[   1 + \frac{a(t)^2 H(t)^2}{k \, c_\psi^2(t)} q(t) \right] w(y)  = 0,
\end{equation}
where
\begin{align}
	q(t) & = -2 + \epsilon_{H1}(t)  + \frac{3}{2} \epsilon_{Z1}(t) + \frac{\epsilon_{1c}(t)}{2} \nonumber \\
    &\qquad -\frac{\epsilon_{1H} (t) \epsilon_{1Z} (t)}{2}  - \frac{\epsilon_{1Z} (t)^2 }{4} - \frac{ \epsilon_{1Z} (t) \epsilon_{2Z} (t)   }{2}  - \frac{ \epsilon_{1c} (t) \epsilon_{1H} (t)   }{2}-
	\frac{ \epsilon_{1c}(t) \epsilon_{2c}(t)}{2}      + \frac{\epsilon_{1c}^2(t)}{4}  .
\end{align}
The above expression is exact in $\epsilon_{1H}(t)$, $\epsilon_{1Z}(t)$, $\epsilon_{1c}(t)$, etc. As we define the dimensionless variable $y \equiv - k \tau$, we can write each flow parameter in terms of $y$ by implementing a logarithmic expansion around the peculiar time $\tau_k\equiv -1/k$, i.e., around $y_k = 1$,
	\begin{align}
			\label{eq:LogExpansion}
		\rho (y)
			&= \rho_k \left[ 1 + \Big(\epsilon_{1\rho k}  + \epsilon_{1\rho k} ( \epsilon_{1H k}-\epsilon_{1c k} )  + \epsilon_{1\rho k} [\epsilon_{1H k} (\epsilon_{1H k} - \epsilon_{2H k}) + \epsilon_{1c k} (-2 \epsilon_{1H k} + \epsilon_{2c k}) +\epsilon_{1c k}^2 ] \Big) \ln\left(y\right) \right.\nonumber  \\
			&\quad + \frac{1}{2}\,\Big( \epsilon_{1\rho k} (\epsilon_{1\rho k} + \epsilon_{2\rho k})  + \epsilon_{1\rho k} [-\epsilon_{1c k} (2 \epsilon_{1\rho k} + \epsilon_{2c k} + 2 \epsilon_{2\rho k}) + \epsilon_{1H k} (2 \epsilon_{1\rho k} + \epsilon_{2H k} + 2 \epsilon_{2\rho k})] \Big) \ln\left(y\right)^2 \nonumber \\
			&\quad \left. + \frac{1}{6} \, \epsilon_{1\rho k} \Big(\epsilon_{1\rho k}^2 + 3 \epsilon_{1\rho k} \epsilon_{2\rho k} + \epsilon_{2\rho k} (\epsilon_{2\rho k} + \epsilon_{3\rho k})\Big) \ln\left(y\right)^3 \right] + \order{\epsilon^4}\,, \nonumber \\
			\epsilon_{1\rho}(y) 
				&=\epsilon_{1\rho k}  + \Big(\epsilon_{1\rho k} \epsilon_{2\rho k} + (-\epsilon_{1c k} + \epsilon_{1H k}) \epsilon_{1\rho k} \epsilon_{2\rho k} \Big) \ln\left(y\right) + \Big( \frac{1}{2} \epsilon_{1\rho k} \epsilon_{2\rho k} (\epsilon_{2\rho k} + \epsilon_{3\rho k})  \Big) \ln\left(y\right)^2 + \order{\epsilon^4}\,, \nonumber \\
				\epsilon_{2\rho}(x) &= \epsilon_{2\rho k} + \epsilon_{2\rho k} \epsilon_{3\rho k} \ln\left(y\right) + \order{\epsilon^3}\,, \nonumber \\
				\epsilon_{3\rho}(x) &= \epsilon_{3\rho k} + \order{\epsilon^2}\,.
	\end{align}
After expressing the quantities in terms of $y=-k\, \tau$, and plugging this logarithmic expansion, the exact equation \eqref{eq:EoM-v(x)-aH} reduces to the mode equation \eqref{eq:EOM} up to N3LO, with the function $g(y)\; =\; g_{1k} + g_{2k}\, \ln(y)\,+ g_{3k}\, \ln(y)^2 + \,\order{\epsilon^3}$, where the coefficients are
\begin{align}
 \label{eq:g's-full-expression}
		g_{1k} &=3 \epsilon_{H1k} - \frac{3  \epsilon_{Z1k} }{2} - \frac{9  \epsilon_{c1k} }{2}  \nonumber \\
		&\qquad +  4 \epsilon_{1Hk}^2 - \frac{5 \epsilon_{1Hk}   \epsilon_{1Zk}}{2} + \frac{ \epsilon_{1Zk}^2}{4} 
		- 4  \epsilon_{1Hk}  \epsilon_{2Hk} + \frac{ \epsilon_{1Zk}  \epsilon_{2Zk}}{2} - \frac{21  \epsilon_{1Hk}  \epsilon_{1ck}}{2} + 3  \epsilon_{1Zk}  \epsilon_{1ck} + \frac{27  \epsilon_{1ck}^2}{4} + \frac{9}{2}  \epsilon_{1ck}  \epsilon_{2ck}		\nonumber \\
		& \qquad  \quad +5 \epsilon_{1Hk}^3 + 4 \epsilon_{1Hk} \epsilon_{2Hk} \epsilon_{3Hk}	- 14 \epsilon_{1Hk}^2 \epsilon_{2Hk}  + 4 \epsilon_{1Hk} \epsilon_{2Hk}^2	 - \frac{7}{2} \epsilon_{1Hk}^2 \epsilon_{1Zk} + \frac{1}{2} \epsilon_{1Hk} \epsilon_{1Zk}^2 + 3 \epsilon_{1Hk} \epsilon_{1Zk} \epsilon_{2Hk} \nonumber \\
		& \qquad \quad + \epsilon_{1Hk} \epsilon_{1Zk} \epsilon_{2Zk}- \frac{1}{2} \epsilon_{1ck} \epsilon_{1Zk}^2 - 18 \epsilon_{1ck}^2 \epsilon_{2ck} + 15 \epsilon_{1ck} \epsilon_{1Hk} \epsilon_{2ck} - 3 \epsilon_{1ck} \epsilon_{1Zk} \epsilon_{2ck} - 4 \epsilon_{1ck} \epsilon_{2ck}^2 + 17 \epsilon_{1ck} \epsilon_{1Hk} \epsilon_{2Hk}  \nonumber \\
		& \qquad   \quad  - \epsilon_{1ck} \epsilon_{1Zk} \epsilon_{2Zk}  - 4 \epsilon_{1ck} \epsilon_{2ck} \epsilon_{3ck}  -9 \epsilon_{1ck}^3 + \frac{45}{2} \epsilon_{1ck}^2 \epsilon_{1Hk} - \frac{37}{2} \epsilon_{1ck} \epsilon_{1Hk}^2  - \frac{9}{2} \epsilon_{1ck}^2 \epsilon_{1Zk} + 8 \epsilon_{1ck} \epsilon_{1Hk} \epsilon_{1Zk}\,, \nonumber \\
		g_{2k} &= 3  \epsilon_{1Hk}  \epsilon_{2Hk} - \frac{3}{2}  \epsilon_{1Zk}  \epsilon_{2Zk} - \frac{9}{2}  \epsilon_{1ck}  \epsilon_{2ck} \nonumber \\
		& \qquad 
			- 4 \epsilon_{1Hk} \epsilon_{2Hk} \epsilon_{3Hk}+ 11 \epsilon_{1Hk}^2 \epsilon_{2Hk}  - 4 \epsilon_{1Hk} \epsilon_{2Hk}^2 + \frac{1}{2} \epsilon_{1Zk} \epsilon_{2Zk} \epsilon_{3Zk} - \frac{5}{2}  \epsilon_{1Hk} \epsilon_{1Zk} \epsilon_{2Hk}  + \frac{9}{2} \epsilon_{1Zk} \epsilon_{2Zk} \epsilon_{1ck} \nonumber \\
			&\qquad - 4\epsilon_{1Zk} \epsilon_{1Hk}  \epsilon_{2Zk} + \frac{1}{2} \epsilon_{1Zk}^2\epsilon_{2Zk}  + \frac{1}{2} \epsilon_{1Zk} \epsilon_{2Zk}^2  + 3 \epsilon_{1ck} \epsilon_{1Zk} \epsilon_{2ck} + \frac{9}{2} \epsilon_{1ck}  \epsilon_{2ck}^2- \frac{27}{2} \epsilon_{1Hk} \epsilon_{2Hk} \epsilon_{1ck}  \nonumber \\
			& \qquad + \frac{9}{2} \epsilon_{1ck} \epsilon_{2ck} \epsilon_{3ck}  +18 \epsilon_{1ck}^2\epsilon_{2ck}  - 15 \epsilon_{1ck} \epsilon_{1Hk} \epsilon_{2ck} \,, \nonumber \\
		g_{3k} &=  \frac{3 \epsilon_{1Hk} \epsilon_{2Hk} \epsilon_{3Hk}}{2} + \frac{3 \epsilon_{1Hk} \epsilon_{2Hk}^2}{2}  - \frac{3 \epsilon_{1Zk} \epsilon_{2Zk} \epsilon_{3Zk}}{4} - \frac{3 \epsilon_{1Zk} \epsilon_{2Zk}^2}{4} - \frac{9 \epsilon_{1ck} \epsilon_{2ck} \epsilon_{3ck}}{4} -\frac{9 \epsilon_{1ck} \epsilon_{2ck}^2}{4} \,.
 	\end{align}

As discussed in \cite{Auclair2022,Bianchi:2024qyp}, the primordial power spectrum is proportional to the term $|y\, w(y)|^2$. Using the Green's function method, we find
 \begin{align}
		\big| y\,  w(y) \big|^2  &= 1 - \frac{2C g_{1 k}}{3} - \frac{4g_{1 k}^2}{9} + \frac{2C g_{1 k}^2}{27} + \frac{2C^2 g_{1 k}^2}{9} + \frac{8g_{1 k}^3}{27} + \frac{68C g_{1 k}^3}{243} - \frac{4C^2 g_{1 k}^3}{81}  \nonumber \\
&\quad - \frac{4C^3 g_{1 k}^3}{81} - \frac{2C g_{2 k}}{9} + \frac{C^2 g_{2 k}}{3} - \frac{8g_1 g_2}{27} + \frac{80C g_{1 k} g_{2 k}}{81} + \frac{2}{27}C^2 g_{1 k} g_{2 k} - \frac{2}{9}C^3 g_{1 k} g_{2 k}  \nonumber\\
&\quad + \frac{g_{1 k}^2 \pi^2}{18} - \frac{g_{1 k}^3 \pi^2}{81} - \frac{1}{27}C g_{1 k}^3 \pi^2 - \frac{g_{2 k} \pi^2}{36} + \frac{7}{162}g_{1 k} g_{2 k} \pi^2 - \frac{5}{54}C g_{1 k} g_{2 k} \pi^2 - \frac{g_{3 k} \pi^2}{54} + \frac{1}{18}C g_{3 k} \pi^2  \nonumber \\
&\quad + \frac{8g_{3 k}}{9} - \frac{4C g_{3 k}}{27} + \frac{2C^2 g_{3 k}}{9} - \frac{2C^3 g_{3 k}}{9} \nonumber  \\
&\qquad + \ln(y) \left(-\frac{2g_{1 k}}{3} + \frac{2g_{1 k}^2}{27} + \frac{4C g_{1 k}^2}{9} + \frac{68g_{1 k}^3}{243} - \frac{8C g_{1 k}^3}{81} - \frac{4C^2 g_{1 k}^3}{27} - \frac{2g_{2 k}}{9} + \frac{8g_{1 k} g_{2 k}}{81}\right.  \nonumber \\
&\qquad + \left.\frac{8C g_{1 k} g_{2 k}}{27} - \frac{2}{9}C^2 g_{1 k} g_{2 k} - \frac{4g_{3 k}}{27} - \frac{g_{1 k}^3 \pi^2}{27} + \frac{1}{54}g_{1 k} g_{2 k} \pi^2\right) - \frac{14}{81}g_{1 k}^3 \zeta(3) - \frac{4}{9}g_{3 k} \zeta(3) \nonumber \\
& \quad\qquad+ \left(\frac{2g_{1 k}^2}{9} - \frac{4g_{1 k}^3}{81} - \frac{4C g_{1 k}^3}{27} - \frac{g_{2 k}}{3} + \frac{2g_{1 k} g_{2 k}}{9}  + \frac{2C g_{1 k} g_{2 k}}{9} - \frac{2g_{3 k}}{9}\right) \ln(y)^2 \nonumber \\
& \quad\qquad - \left(\frac{4g_{1 k}^3}{81} - \frac{2g_{1 k} g_{2 k}}{9} + \frac{2g_{3 k}}{9}\right) \ln(y)^3
\,, \label{eq:asymptotic}
\end{align}
where $C = \gamma_E + \ln(2) - 2 \simeq -0.730$. To find an expression of $\mathcal{P}^{(\psi)}(k)$ fully expanded around a pivot scale $k_\ast$ defined as $k_\ast \tau_\ast = -1$, first note that one can perform a new logarithmic expansion, similar to \eqref{eq:LogExpansion}, but around a variable $y_\ast=k/k_\ast$, i.e., $\rho(y)=\rho_\ast (1+\epsilon_{1\rho\ast}\, \ln(y/y_\ast) + \cdots )$. Then, note that $\ln(y/y_\ast)=-\ln(k/k_\ast)$. Replacing this new expansion into $Z(y(t))$, $H(y(t))$, $c(y(t))$, and \eqref{eq:asymptotic}, one find the final expression for the primordial power spectrum \eqref{eq:Power-BunchDavies}. The coefficients are exactly the same as the ones in Tables III, IV, V, and VI of \cite{Bianchi:2024qyp}, with the only small difference that our $p_{0\ast}$ in \eqref{eq:Power-BunchDavies} is reported in \cite{Bianchi:2024qyp} as $1+p_{0\ast}$.

\section{An accurate computation of $N_{\odot}$}
\label{app:Computation_N_odot}

We present an accurate computation of the number of e-foldings $N_\odot$ in $R+\alpha\, R^2 - \beta\, W^2$ inflation. Note that, for $\beta=0$, the result also provides an accurate calculation of $N_\ast$ for Starobinsky inflation in the Jordan frame. The number of e-foldings $N_{\odot}$ accounts for the cosmological expansion of the Universe from the time $t_{\odot}$ at which the pivot scale $k_\odot$ freezes until the time $t_{\mathrm{end}}$  when inflation ends, defined as $\epsilon_{1H}(t_{\mathrm{end}}) = 1$. First, the modified Friedman equation  \eqref{eq:Friedmann_Modified} gives us
\begin{equation}
	\label{eq:epsilon_2H_end}
\epsilon_{2H}(t_{\mathrm{end}}) = \frac{3}{2} - \frac{1}{12\, \alpha \, H_{\mathrm{end}}^2} \, .
\end{equation}
To give a precise evaluation of $H_{\mathrm{end}}$ we should not use  \eqref{eq:H-to-eps}, as the expansion cannot be trusted for $\epsilon_{1H}\approx 1$. A better strategy is to find its Padé approximant, which is given by
\begin{equation}
[2/2]_{H(t)} = \frac{5472 + 8496 \, \epsilon_{1H}(t) -385 \, \epsilon_{1H}(t)^2   }{144\, (228 + 373\, \epsilon_{1H}(t)) \, \sqrt{\alpha\, \epsilon_{1H}(t)}  } \, .
\end{equation}
The Taylor expansion of the above expression for small $\epsilon_{1H}(t)$ matches the expansion \eqref{eq:H-to-eps} up to order $\order{\epsilon_{1H}^4}$. Evaluating on $\epsilon_{1H}(t_{\mathrm{end}}) =1$ allows us to write 
\begin{equation}
\label{eq:H_end}
H_{\mathrm{end}} \approx \frac{13583}{86544} \, \frac{1}{\sqrt{\alpha}} \approx  \frac{1}{\sqrt{40.6 \times \alpha}} .
\end{equation}
Note that this value of $H_{\mathrm{end}}$ slightly decreases the leading order scale $1/\sqrt{36 \, \alpha}$, which is consistent with numerical solutions of \eqref{eq:Friedmann_Modified}, as shown in Fig. \ref{fig:starobinsky}. Additionally, replacing $H_{\mathrm{end}}$ into \eqref{eq:epsilon_2H_end} gives us $\epsilon_{2H}(t_{\mathrm{end}}) \approx - 1.883$. One can follow a similar procedure to find a modified Padé approximant for $\epsilon_{2H}(t)$ as a function of $\epsilon_{1H}(t)$.  Let us assume a rational function of the form
\begin{equation}
\epsilon_{2H}(t) = - \frac{a\, \epsilon_{1H}(t)  + b\, \epsilon_{1H}(t)^2 }{1+ c\,  \epsilon_{1H}(t)  + d\,  \epsilon_{1H}(t)^2 + e \, \epsilon_{1H}(t)^3  }\, ,
\end{equation}
and we impose four matching conditions with the expansion \eqref{eq:Epsilon_Expansion_234}, and one last restriction to $\epsilon_{2H}(t_{\mathrm{end}})= -354/188 \approx - 1.883 $. These five conditions give a modified [2/3] Padé approximant of the form,
\begin{equation}
	\label{eq:Pade-epsilon_2H}
[2/3]_{\epsilon_{2H}(t)} =  - \frac{2\, \epsilon_{1H}(t)  + \tfrac{5993}{927} \, \epsilon_{1H}(t)^2 }{1+ \tfrac{3151}{927}\, \epsilon_{1H}(t) + \tfrac{803}{2781}\, \epsilon_{1H}(t)^2 - \tfrac{535}{2781}\, \epsilon_{1H}(t)^3  } \, .
\end{equation}
The amount of cosmological expansion from the time $t_\odot$, associated with the pivot scale $k_\odot \tau_\odot = -1$ until the end of inflation, is given by the e-folding number $N_\odot$, defined as
\begin{equation}
	a(t_{\mathrm{end}})=\ee^{N_\odot}\,a(t_\odot)\,.
\end{equation}
To find $N_\odot$, an analytical expression as the integral over the Hubble flow parameters is available,
\begin{equation}
N_\odot \equiv\int_{t_\odot}^{t_{\textrm{end}}} H(t) \dd{t} = - \int_{\epsilon_{1H\odot}}^{1} \frac{\dd{\epsilon_{1H}}}{ \epsilon_{1H}\;\, \epsilon_{2H}\big(\epsilon_{1H}\big)  } \, ,
\end{equation}
and  after replacing \eqref{eq:Pade-epsilon_2H}, we find
\begin{equation}
N_\odot	=  \frac{1}{2 \, \epsilon_{1H\odot}  } + \frac{1}{12} \ln(\frac{1}{\epsilon_{1H\odot}}) - \frac{19049}{35958} + \frac{535}{17979} \epsilon_{1H\odot} + \frac{4 232 991\, \ln(\tfrac{1854+5993\, \epsilon_{1H\odot}}{7847})}{143 664 196}  \, .
\end{equation}
The expansion for small $\epsilon_{1H\odot}$ gives a final expression 
\begin{equation}
	\label{eq:N_odot_epsilon}
N_\odot = \frac{1}{2 \, \epsilon_{1H\odot}  } + \frac{1}{12} \ln(\frac{1}{\epsilon_{1H\odot}}) - D_0 + \frac{\epsilon_{1H\odot}}{8} +\order{\epsilon_{1H\odot}^2}
\end{equation}
where $D_0 = 19049/35958 + 4232991\, \ln(\tfrac{7847}{1854})/143 664 196 \approx 0.57$. This result estimates the number of e-foldings with percent-level accuracy, as confirmed by direct numerical integration. Moreover, the calculation of the constant $D_0$ represents a refinement of the results in 
\cite{Bianchi:2024qyp}, where it was previously estimated to be $D_0\approx 0.60$ using a different approximation scheme.

It is useful to have a relation expressed uniquely in terms of observables, in particular $n_\mathrm{s}$. To find a relation of the form $N_\odot (n_\mathrm{s})$, we can write an ansatz of the form
\begin{equation}
N_\odot = \frac{a}{1-n_\mathrm{s}} + b \, \ln(\frac{1}{1-n_\mathrm{s}}) + c\, (1-n_{\mathrm{s}}) + d + \order{(1-n_\mathrm{s})^2} \, .
\end{equation}
Then, we replace the expansion $\epsilon_{1H\odot}$ to $n_\mathrm{s}$, and by matching the coefficients of \eqref{eq:N_odot_epsilon}, we find
\begin{equation}
	\label{eq:N_odot_ns}
N_\odot = \frac{2}{1-n_{\mathrm{s}}}  + \frac{1}{12} \ln(\frac{1}{1-n_{\mathrm{s}}}) - (c_0+\tfrac{\beta}{48\,\alpha}) (1-n_{\mathrm{s}})- d_0 + \order{(1-n_\mathrm{s})^2}\, ,
\end{equation}
where $c_0=7\pi^2/24-C/4-707/288 \approx 0.606$, and $d_0 = (7+\ln(2)+6 \, C - 6 D_0 )/6 \approx 0.02$.  One can also invert \eqref{eq:N_odot_ns} and find,
\begin{equation}
n_{\mathrm{s}} = 1- \frac{2}{N_\odot} + \frac{\theta_1}{N_\odot^2} - \frac{1}{6} \, \frac{\ln(N_\odot)}{N_\odot^2}  + \frac{\theta_2\, \ln(N_\odot)}{N_\odot^3}  -\frac{1}{72} \frac{\ln(N_\odot)^2}{N_\odot^3} + \frac{\theta_3+\tfrac{\beta}{12\, \alpha}}{N_\odot^3}+ \order{\frac{1}{N_\odot^4}} \, ,
\end{equation}
where $\theta_1 = \ln(2)/6+2\,d_0\approx 0.155$, $\theta_2 = (1+24\, d_0 + 2\ln(2))/72\approx 0.040$, and 
$\theta_3 = (-12\, d_0 - 144 \, d_0^2 + 288\, c_0 - \ln(2)- 24\, d_0 \ln(2)- \ln(2)^2 )/72 \approx 2.399$. This expression refines a previous calculation found in \cite{Bianchi:2024qyp}.

It is important to remark that a dependence on $\beta/\alpha$ enters only as a subleading correction at order $\order{N_\odot^{-3}}$. Consequently, we drop the $\odot$ label and use $N_\ast$, with the associated pivot $k_\ast = a_\ast H_\ast$, to facilitate the comparison with observations and other works in the literature. Note also that $N_\ast$ corresponds to the number of e-foldings in the Jordan frame, while in the literature it is usually reported as $N_\ast^{(E)}$, i.e., the number of e-foldings in the Einstein frame. Both quantities are related via Eq. \eqref{eq:Jordan-Einstein}.

%\end{widetext}

\vfill
\twocolumngrid

 %\bibliographystyle{JHEP}
 %\bibliography{bib-inflation}

\begin{thebibliography}{100}

\bibitem{Brout1978}
R.~Brout, F.~Englert and E.~Gunzig, \emph{The creation of the universe as a
  quantum phenomenon.},
  \href{https://doi.org/10.1016/0003-4916(78)90176-8}{\emph{Annals of Physics}
  {\bfseries 115} (1978) 78}.

\bibitem{Starobinsky1979}
A.A.~Starobinsky, \emph{Spectrum of relict gravitational radiation and the
  early state of the universe},
  {\href{http://jetpletters.ru/ps/1370/article_20738.shtml}{\emph{JETP Lett.} {\bfseries 30} (1979) 682}}.

\bibitem{Starobinsky1980}
A.A.~{Starobinsky}, \emph{{A new type of isotropic cosmological models without
  singularity}},
  \href{https://doi.org/10.1016/0370-2693(80)90670-X}{\emph{Physics Letters B}
  {\bfseries 91} (1980) 99}.

\bibitem{Mukhanov1981}
V.F.~Mukhanov and G.V.~Chibisov, \emph{{Quantum Fluctuations and a Nonsingular
  Universe}},
  {\href{http://jetpletters.ru/ps/1510/article_23079.shtml}{\emph{JETP Lett.}
  {\bfseries 33} (1981) 532}}.

\bibitem{Sato:1981ds}
K.~Sato, \emph{{Cosmological Baryon Number Domain Structure and the First Order
  Phase Transition of a Vacuum}},
  \href{https://doi.org/10.1016/0370-2693(81)90805-4}{\emph{Phys. Lett. B}
  {\bfseries 99} (1981) 66}.

\bibitem{Sato:1981qmu}
K.~Sato, \emph{{First-order phase transition of a vacuum and the expansion of
  the Universe}}, \href{https://doi.org/10.1093/mnras/195.3.467}{\emph{Mon.
  Not. Roy. Astron. Soc.} {\bfseries 195} (1981) 467}.

\bibitem{Guth1981}
A.H.~{Guth}, \emph{{Inflationary universe: A possible solution to the horizon
  and flatness problems}},
  \href{https://doi.org/10.1103/PhysRevD.23.347}{\emph{\prd} {\bfseries 23}
  (1981) 347}.

\bibitem{Linde1982}
A.D.~{Linde}, \emph{{A new inflationary universe scenario: A possible solution
  of the horizon, flatness, homogeneity, isotropy and primordial monopole
  problems}}, \href{https://doi.org/10.1016/0370-2693(82)91219-9}{\emph{Physics
  Letters B} {\bfseries 108} (1982) 389}.

\bibitem{Albrecht1982}
A.~{Albrecht} and P.J.~{Steinhardt}, \emph{{Cosmology for Grand Unified
  Theories with Radiatively Induced Symmetry Breaking}},
  \href{https://doi.org/10.1103/PhysRevLett.48.1220}{\emph{\prl} {\bfseries 48}
  (1982) 1220}.

\bibitem{Guth1982}
A.H.~Guth and S.Y.~Pi, \emph{Fluctuations in the new inflationary universe},
  \href{https://doi.org/10.1103/PhysRevLett.49.1110}{\emph{Phys. Rev. Lett.}
  {\bfseries 49} (1982) 1110}.

\bibitem{Hawking1982}
S.W.~Hawking, \emph{The development of irregularities in a single bubble
  inflationary universe},
  \href{https://doi.org/10.1016/0370-2693(82)90373-2}{\emph{Phys. Lett. B}
  {\bfseries 115} (1982) 295}.

\bibitem{Linde1983}
A.D.~Linde, \emph{Chaotic inflation},
  \href{https://doi.org/10.1016/0370-2693(83)90837-7}{\emph{Phys. Lett. B}
  {\bfseries 129} (1983) 177}.

\bibitem{Calabrese:2013jyk}
E.~Calabrese et~al., \emph{{Cosmological parameters from pre-planck cosmic
  microwave background measurements}},
  \href{https://doi.org/10.1103/PhysRevD.87.103012}{\emph{Phys. Rev. D}
  {\bfseries 87} (2013) 103012}
  [\href{https://arxiv.org/abs/1302.1841}{{\ttfamily 1302.1841}}].

\bibitem{Planck:2018jri}
{\scshape Planck} collaboration, \emph{{Planck 2018 results. X. Constraints on
  inflation}}, \href{https://doi.org/10.1051/0004-6361/201833887}{\emph{Astron.
  Astrophys.} {\bfseries 641} (2020) A10}
  [\href{https://arxiv.org/abs/1807.06211}{{\ttfamily 1807.06211}}].

\bibitem{KeckCollaboration2021}
{Keck Collaboration}, P.A.R.~Ade, Z.~Ahmed et~al., \emph{Bicep / keck xiii:
  Improved constraints on primordial gravitational waves using planck, wmap,
  and bicep/keck observations through the 2018 observing season},
  \href{https://doi.org/10.1103/physrevlett.127.151301}{\emph{Phys. Rev. Lett.
  127, 151301 (2021)} {\bfseries 127} (2021) 151301}
  [\href{https://arxiv.org/abs/2110.00483}{{\ttfamily 2110.00483}}].

\bibitem{ACT:2020frw}
{\scshape ACT} collaboration, \emph{{The Atacama Cosmology Telescope: a
  measurement of the Cosmic Microwave Background power spectra at 98 and 150
  GHz}}, \href{https://doi.org/10.1088/1475-7516/2020/12/045}{\emph{JCAP}
  {\bfseries 12} (2020) 045}
  [\href{https://arxiv.org/abs/2007.07289}{{\ttfamily 2007.07289}}].

\bibitem{ACT:2025tim}
{\scshape ACT} collaboration, \emph{{The Atacama Cosmology Telescope: DR6
  Constraints on Extended Cosmological Models}},
  [\href{https://arxiv.org/abs/2503.14454}{{\ttfamily 2503.14454}}].

\bibitem{Martin:2013tda}
J.~Martin, C.~Ringeval and V.~Vennin, \emph{{Encyclop\ae{}dia Inflationaris}:
  {Opiparous Edition}},
  \href{https://doi.org/10.1016/j.dark.2024.101653}{\emph{Phys. Dark Univ.}
  {\bfseries 5-6} (2014) 75} [\href{https://arxiv.org/abs/1303.3787}{{\ttfamily
  1303.3787}}].

\bibitem{Kallosh:2025ijd}
R.~Kallosh and A.~Linde, \emph{{On the present status of inflationary
  cosmology}}, \href{https://doi.org/10.1007/s10714-025-03470-6}{\emph{Gen.
  Rel. Grav.} {\bfseries 57} (2025) 135}
  [\href{https://arxiv.org/abs/2505.13646}{{\ttfamily 2505.13646}}].

\bibitem{Martin:2024qnn}
J.~Martin, C.~Ringeval and V.~Vennin, \emph{{Cosmic Inflation at the
  crossroads}},
  \href{https://doi.org/10.1088/1475-7516/2024/07/087}{\emph{JCAP} {\bfseries
  07} (2024) 087} [\href{https://arxiv.org/abs/2404.10647}{{\ttfamily
  2404.10647}}].

\bibitem{Martin:2010hh}
J.~Martin, C.~Ringeval and R.~Trotta, \emph{{Hunting Down the Best Model of
  Inflation with Bayesian Evidence}},
  \href{https://doi.org/10.1103/PhysRevD.83.063524}{\emph{Phys. Rev. D}
  {\bfseries 83} (2011) 063524}
  [\href{https://arxiv.org/abs/1009.4157}{{\ttfamily 1009.4157}}].

\bibitem{Ballardini:2024ado}
M.~Ballardini, \emph{{Chasing cosmic inflation: constraints for inflationary
  models and reheating insights}},
  \href{https://doi.org/10.1088/1475-7516/2025/01/116}{\emph{JCAP} {\bfseries
  01} (2025) 116} [\href{https://arxiv.org/abs/2408.03321}{{\ttfamily
  2408.03321}}].

\bibitem{Sorensen:2024ezb}
C.T.G.~S\o{}rensen, S.~Hannestad, A.~Nygaard and T.~Tram, \emph{{Calculating
  Bayesian evidence for inflationary models using connect}},
  \href{https://doi.org/10.1088/1475-7516/2025/03/043}{\emph{JCAP} {\bfseries
  03} (2025) 043} [\href{https://arxiv.org/abs/2406.03968}{{\ttfamily
  2406.03968}}].

\bibitem{tHooft:1974toh}
G.~'t~Hooft and M.J.G.~Veltman, \emph{{One loop divergencies in the theory of
  gravitation}}, {\emph{Ann. Inst. H. Poincare A Phys. Theor.} {\bfseries 20}
  (1974) 69}.

\bibitem{Stelle1977}
K.S.~Stelle, \emph{{Renormalization of Higher Derivative Quantum Gravity}},
  \href{https://doi.org/10.1103/PhysRevD.16.953}{\emph{Phys. Rev. D} {\bfseries
  16} (1977) 953}.

\bibitem{Stelle1978}
K.S.~Stelle, \emph{{Classical Gravity with Higher Derivatives}},
  \href{https://doi.org/10.1007/BF00760427}{\emph{Gen. Rel. Grav.} {\bfseries
  9} (1978) 353}.

\bibitem{Birrell:1982ix}
N.D.~Birrell and P.C.W.~Davies, \emph{{Quantum Fields in Curved Space}},
  Cambridge Monographs on Mathematical Physics, Cambridge University Press,
  Cambridge, UK (1982),
  \href{https://doi.org/10.1017/CBO9780511622632}{10.1017/CBO9780511622632}.

\bibitem{Donoghue:1994dn}
J.F.~Donoghue, \emph{{General relativity as an effective field theory: The
  leading quantum corrections}},
  \href{https://doi.org/10.1103/PhysRevD.50.3874}{\emph{Phys. Rev. D}
  {\bfseries 50} (1994) 3874}
  [\href{https://arxiv.org/abs/gr-qc/9405057}{{\ttfamily gr-qc/9405057}}].

\bibitem{Burgess:2003jk}
C.P.~Burgess, \emph{{Quantum gravity in everyday life: General relativity as an
  effective field theory}},
  \href{https://doi.org/10.12942/lrr-2004-5}{\emph{Living Rev. Rel.} {\bfseries
  7} (2004) 5} [\href{https://arxiv.org/abs/gr-qc/0311082}{{\ttfamily
  gr-qc/0311082}}].

\bibitem{Buchbinder:2017lnd}
I.L.~Buchbinder, S.D.~Odintsov and I.L.~Shapiro, \emph{{Effective Action in
  Quantum Gravity}}, Routledge (10, 2017),
  \href{https://doi.org/10.1201/9780203758922}{10.1201/9780203758922}.

\bibitem{Donoghue:2022eay}
J.F.~Donoghue, \emph{{Quantum General Relativity and Effective Field Theory}},
  (2023), \href{https://doi.org/10.1007/978-981-19-3079-9_1-1}{DOI}
  [\href{https://arxiv.org/abs/2211.09902}{{\ttfamily 2211.09902}}].

\bibitem{Berkin:1991nb}
A.L.~Berkin, \emph{{Contribution of the Weyl tensor to $R^2$ inflation}},
  \href{https://doi.org/10.1103/PhysRevD.44.1020}{\emph{Phys. Rev. D}
  {\bfseries 44} (1991) 1020}.

\bibitem{Weinberg:2008}
S.~Weinberg, \emph{{Effective Field Theory for Inflation}},
  \href{https://doi.org/10.1103/PhysRevD.77.123541}{\emph{Phys. Rev. D}
  {\bfseries 77} (2008) 123541}
  [\href{https://arxiv.org/abs/0804.4291}{{\ttfamily 0804.4291}}].

\bibitem{Weinberg:2009wa}
S.~Weinberg, \emph{{Asymptotically Safe Inflation}},
  \href{https://doi.org/10.1103/PhysRevD.81.083535}{\emph{Phys. Rev. D}
  {\bfseries 81} (2010) 083535}
  [\href{https://arxiv.org/abs/0911.3165}{{\ttfamily 0911.3165}}].

\bibitem{Clunan:2009er}
T.~Clunan and M.~Sasaki, \emph{{Tensor ghosts in the inflationary cosmology}},
  \href{https://doi.org/10.1088/0264-9381/27/16/165014}{\emph{Class. Quant.
  Grav.} {\bfseries 27} (2010) 165014}
  [\href{https://arxiv.org/abs/0907.3868}{{\ttfamily 0907.3868}}].

\bibitem{Deruelle:2010kf}
N.~Deruelle, M.~Sasaki, Y.~Sendouda and A.~Youssef, \emph{{Inflation with a
  Weyl term, or ghosts at work}},
  \href{https://doi.org/10.1088/1475-7516/2011/03/040}{\emph{JCAP} {\bfseries
  03} (2011) 040} [\href{https://arxiv.org/abs/1012.5202}{{\ttfamily
  1012.5202}}].

\bibitem{Deruelle:2012xv}
N.~Deruelle, M.~Sasaki, Y.~Sendouda and A.~Youssef, \emph{{Lorentz-violating vs
  ghost gravitons: the example of Weyl gravity}},
  \href{https://doi.org/10.1007/JHEP09(2012)009}{\emph{JHEP} {\bfseries 09}
  (2012) 009} [\href{https://arxiv.org/abs/1202.3131}{{\ttfamily 1202.3131}}].

\bibitem{Myrzakulov:2014hca}
R.~Myrzakulov, S.~Odintsov and L.~Sebastiani, \emph{{Inflationary universe from
  higher-derivative quantum gravity}},
  \href{https://doi.org/10.1103/PhysRevD.91.083529}{\emph{Phys. Rev. D}
  {\bfseries 91} (2015) 083529}
  [\href{https://arxiv.org/abs/1412.1073}{{\ttfamily 1412.1073}}].

\bibitem{DeLaurentis:2015fea}
M.~De~Laurentis, M.~Paolella and S.~Capozziello, \emph{{Cosmological inflation
  in $F(R,\mathcal{G})$ gravity}},
  \href{https://doi.org/10.1103/PhysRevD.91.083531}{\emph{Phys. Rev. D}
  {\bfseries 91} (2015) 083531}
  [\href{https://arxiv.org/abs/1503.04659}{{\ttfamily 1503.04659}}].

\bibitem{Baumann2015}
D.~Baumann, H.~Lee and G.L.~Pimentel, \emph{High-scale inflation and the tensor
  tilt}, \href{https://doi.org/10.1007/jhep01(2016)101}{\emph{Journal of High
  Energy Physics} {\bfseries 2016} (2015) }
  [\href{https://arxiv.org/abs/1507.07250}{{\ttfamily 1507.07250}}].

\bibitem{Myung:2014jha}
Y.S.~Myung and T.~Moon, \emph{{Primordial massive gravitational waves from
  Einstein-Chern-Simons-Weyl gravity}},
  \href{https://doi.org/10.1088/1475-7516/2014/08/061}{\emph{JCAP} {\bfseries
  08} (2014) 061} [\href{https://arxiv.org/abs/1406.4367}{{\ttfamily
  1406.4367}}].

\bibitem{Myung:2015vya}
Y.S.~Myung and T.~Moon, \emph{{Scale-invariant tensor spectrum from conformal
  gravity}}, \href{https://doi.org/10.1142/S0217732315501722}{\emph{Mod. Phys.
  Lett. A} {\bfseries 30} (2015) 1550172}
  [\href{https://arxiv.org/abs/1501.01749}{{\ttfamily 1501.01749}}].

\bibitem{Ivanov:2016hcm}
M.M.~Ivanov and A.A.~Tokareva, \emph{{Cosmology with a light ghost}},
  \href{https://doi.org/10.1088/1475-7516/2016/12/018}{\emph{JCAP} {\bfseries
  12} (2016) 018} [\href{https://arxiv.org/abs/1610.05330}{{\ttfamily
  1610.05330}}].

\bibitem{Salvio:2017xul}
A.~Salvio, \emph{{Inflationary Perturbations in No-Scale Theories}},
  \href{https://doi.org/10.1140/epjc/s10052-017-4825-6}{\emph{Eur. Phys. J. C}
  {\bfseries 77} (2017) 267}
  [\href{https://arxiv.org/abs/1703.08012}{{\ttfamily 1703.08012}}].

\bibitem{Cano:2020oaa}
P.A.~Cano, K.~Fransen and T.~Hertog, \emph{{Novel higher-curvature variations
  of $R^2$ inflation}},
  \href{https://doi.org/10.1103/PhysRevD.103.103531}{\emph{Phys. Rev. D}
  {\bfseries 103} (2021) 103531}
  [\href{https://arxiv.org/abs/2011.13933}{{\ttfamily 2011.13933}}].

\bibitem{Anselmi2020}
D.~Anselmi, E.~Bianchi and M.~Piva, \emph{Predictions of quantum gravity in
  inflationary cosmology: effects of the weyl-squared term},
  \href{https://doi.org/10.1007/jhep07(2020)211}{\emph{J. High Energy Phys. 07
  (2020) 211} {\bfseries 2020} (2020) }
  [\href{https://arxiv.org/abs/2005.10293}{{\ttfamily 2005.10293}}].

\bibitem{Giare:2020plo}
W.~Giar\`e, F.~Renzi and A.~Melchiorri, \emph{{Higher-Curvature Corrections and
  Tensor Modes}},
  \href{https://doi.org/10.1103/PhysRevD.103.043515}{\emph{Phys. Rev. D}
  {\bfseries 103} (2021) 043515}
  [\href{https://arxiv.org/abs/2012.00527}{{\ttfamily 2012.00527}}].

\bibitem{Kaczmarek:2021psy}
A.Z.~Kaczmarek and D.~Szcz\k{e}\'sniak, \emph{{Cosmology in the mimetic
  higher-curvature $f(R,R_{\mu \nu }R^{\mu \nu })$ gravity}},
  \href{https://doi.org/10.1038/s41598-021-97907-y}{\emph{Sci. Rep.} {\bfseries
  11} (2021) 18363} [\href{https://arxiv.org/abs/2105.05050}{{\ttfamily
  2105.05050}}].

\bibitem{Gialamas:2021enw}
I.D.~Gialamas, A.~Karam, T.D.~Pappas and V.C.~Spanos, \emph{{Scale-invariant
  quadratic gravity and inflation in the Palatini formalism}},
  \href{https://doi.org/10.1103/PhysRevD.104.023521}{\emph{Phys. Rev. D}
  {\bfseries 104} (2021) 023521}
  [\href{https://arxiv.org/abs/2104.04550}{{\ttfamily 2104.04550}}].

\bibitem{Salvio:2022mld}
A.~Salvio, \emph{{BICEP/Keck data and quadratic gravity}},
  \href{https://doi.org/10.1088/1475-7516/2022/09/027}{\emph{JCAP} {\bfseries
  09} (2022) 027} [\href{https://arxiv.org/abs/2202.00684}{{\ttfamily
  2202.00684}}].

\bibitem{Koshelev:2022olc}
A.S.~Koshelev, K.S.~Kumar and A.A.~Starobinsky, \emph{{Generalized non-local
  R$^{2}$-like inflation}},
  \href{https://doi.org/10.1007/JHEP07(2023)146}{\emph{JHEP} {\bfseries 07}
  (2023) 146} [\href{https://arxiv.org/abs/2209.02515}{{\ttfamily
  2209.02515}}].

\bibitem{Kubo:2022dlx}
J.~Kubo, J.~Kuntz, J.~Rezacek and P.~Saake, \emph{{Inflation with massive
  spin-2 ghosts}},
  \href{https://doi.org/10.1088/1475-7516/2022/11/049}{\emph{JCAP} {\bfseries
  11} (2022) 049} [\href{https://arxiv.org/abs/2207.14329}{{\ttfamily
  2207.14329}}].

\bibitem{DeFelice:2023psw}
A.~De~Felice, R.~Kawaguchi, K.~Mizui and S.~Tsujikawa, \emph{{Starobinsky
  inflation with a quadratic Weyl tensor}},
  \href{https://doi.org/10.1103/PhysRevD.108.123524}{\emph{Phys. Rev. D}
  {\bfseries 108} (2023) 123524}
  [\href{https://arxiv.org/abs/2309.01835}{{\ttfamily 2309.01835}}].

\bibitem{Khodabakhshi:2024med}
H.~Khodabakhshi, M.~Farhang, F.~Shojai and H.~L\"u, \emph{{Cosmology with
  higher-derivative gravities}},
  \href{https://doi.org/10.1103/PhysRevD.110.L061504}{\emph{Phys. Rev. D}
  {\bfseries 110} (2024) L061504}
  [\href{https://arxiv.org/abs/2405.02879}{{\ttfamily 2405.02879}}].

\bibitem{Kubo:2025jla}
J.~Kubo and J.~Kuntz, \emph{{Primordial gravitational waves in quadratic
  gravity}}, \href{https://doi.org/10.1088/1475-7516/2025/05/093}{\emph{JCAP}
  {\bfseries 05} (2025) 093}
  [\href{https://arxiv.org/abs/2502.03543}{{\ttfamily 2502.03543}}].

\bibitem{Modesto:2011kw}
L.~Modesto, \emph{{Super-renormalizable Quantum Gravity}},
  \href{https://doi.org/10.1103/PhysRevD.86.044005}{\emph{Phys. Rev. D}
  {\bfseries 86} (2012) 044005}
  [\href{https://arxiv.org/abs/1107.2403}{{\ttfamily 1107.2403}}].

\bibitem{Koshelev:2016xqb}
A.S.~Koshelev, L.~Modesto, L.~Rachwal and A.A.~Starobinsky, \emph{{Occurrence
  of exact $R^2$ inflation in non-local UV-complete gravity}},
  \href{https://doi.org/10.1007/JHEP11(2016)067}{\emph{JHEP} {\bfseries 11}
  (2016) 067} [\href{https://arxiv.org/abs/1604.03127}{{\ttfamily
  1604.03127}}].

\bibitem{Salvio:2018crh}
A.~Salvio, \emph{{Quadratic Gravity}},
  \href{https://doi.org/10.3389/fphy.2018.00077}{\emph{Front. in Phys.}
  {\bfseries 6} (2018) 77} [\href{https://arxiv.org/abs/1804.09944}{{\ttfamily
  1804.09944}}].

\bibitem{Anselmi:2018ibi}
D.~Anselmi and M.~Piva, \emph{{The Ultraviolet Behavior of Quantum Gravity}},
  \href{https://doi.org/10.1007/JHEP05(2018)027}{\emph{JHEP} {\bfseries 05}
  (2018) 027} [\href{https://arxiv.org/abs/1803.07777}{{\ttfamily
  1803.07777}}].

\bibitem{Donoghue:2021cza}
J.F.~Donoghue and G.~Menezes, \emph{{On quadratic gravity}},
  \href{https://doi.org/10.1393/ncc/i2022-22026-7}{\emph{Nuovo Cim. C}
  {\bfseries 45} (2022) 26} [\href{https://arxiv.org/abs/2112.01974}{{\ttfamily
  2112.01974}}].

\bibitem{Buoninfante:2023ryt}
L.~Buoninfante, \emph{{Massless and partially massless limits in Quadratic
  Gravity}}, \href{https://doi.org/10.1007/JHEP12(2023)111}{\emph{JHEP}
  {\bfseries 12} (2023) 111}
  [\href{https://arxiv.org/abs/2308.11324}{{\ttfamily 2308.11324}}].

\bibitem{Buccio:2024hys}
D.~Buccio, J.F.~Donoghue, G.~Menezes and R.~Percacci, \emph{{Physical Running
  of Couplings in Quadratic Gravity}},
  \href{https://doi.org/10.1103/PhysRevLett.133.021604}{\emph{Phys. Rev. Lett.}
  {\bfseries 133} (2024) 021604}
  [\href{https://arxiv.org/abs/2403.02397}{{\ttfamily 2403.02397}}].

\bibitem{Ashtekar2021}
A.~Ashtekar and E.~Bianchi, \emph{A short review of loop quantum gravity},
  \href{https://doi.org/10.1088/1361-6633/abed91}{\emph{Rep. Prog. Phys. 84,
  042001 (2021)} {\bfseries 84} (2021) 042001}
  [\href{https://arxiv.org/abs/2104.04394}{{\ttfamily 2104.04394}}].

\bibitem{Ashtekar:2011ni}
A.~Ashtekar and P.~Singh, \emph{{Loop Quantum Cosmology: A Status Report}},
  \href{https://doi.org/10.1088/0264-9381/28/21/213001}{\emph{Class. Quant.
  Grav.} {\bfseries 28} (2011) 213001}
  [\href{https://arxiv.org/abs/1108.0893}{{\ttfamily 1108.0893}}].

\bibitem{Bianchi:2010zs}
E.~Bianchi, C.~Rovelli and F.~Vidotto, \emph{{Towards Spinfoam Cosmology}},
  \href{https://doi.org/10.1103/PhysRevD.82.084035}{\emph{Phys. Rev. D}
  {\bfseries 82} (2010) 084035}
  [\href{https://arxiv.org/abs/1003.3483}{{\ttfamily 1003.3483}}].

\bibitem{Gozzini:2019nbo}
F.~Gozzini and F.~Vidotto, \emph{{Primordial Fluctuations From Quantum
  Gravity}}, \href{https://doi.org/10.3389/fspas.2020.629466}{\emph{Front.
  Astron. Astrophys. Cosmol.} {\bfseries 7} (2021) 629466}
  [\href{https://arxiv.org/abs/1906.02211}{{\ttfamily 1906.02211}}].

\bibitem{Gielen:2013kla}
S.~Gielen, D.~Oriti and L.~Sindoni, \emph{{Cosmology from Group Field Theory
  Formalism for Quantum Gravity}},
  \href{https://doi.org/10.1103/PhysRevLett.111.031301}{\emph{Phys. Rev. Lett.}
  {\bfseries 111} (2013) 031301}
  [\href{https://arxiv.org/abs/1303.3576}{{\ttfamily 1303.3576}}].

\bibitem{Dittrich:2021gww}
B.~Dittrich, S.~Gielen and S.~Schander, \emph{{Lorentzian quantum cosmology
  goes simplicial}},
  \href{https://doi.org/10.1088/1361-6382/ac42ad}{\emph{Class. Quant. Grav.}
  {\bfseries 39} (2022) 035012}
  [\href{https://arxiv.org/abs/2109.00875}{{\ttfamily 2109.00875}}].

\bibitem{Agullo:2023rqq}
I.~Agullo, A.~Wang and E.~Wilson-Ewing, \emph{Loop quantum cosmology: Relation
  between theory and observations},  in \emph{Handbook of Quantum Gravity},
  C.~Bambi, L.~Modesto and I.~Shapiro, eds., (Singapore), pp.~1--46, Springer
  Nature Singapore (2023),
  \href{https://doi.org/10.1007/978-981-19-3079-9_103-1}{DOI}
  [\href{https://arxiv.org/abs/2301.10215}{{\ttfamily 2301.10215}}].

\bibitem{Bhabha:1946zz}
H.J.~Bhabha, \emph{{On the Expansibility of Solutions in Powers of the
  Interaction Constants}},
  \href{https://doi.org/10.1103/PhysRev.70.759}{\emph{Phys. Rev.} {\bfseries
  70} (1946) 759}.

\bibitem{Simon:1990ic}
J.Z.~Simon, \emph{{Higher Derivative Lagrangians, Nonlocality, Problems and
  Solutions}}, \href{https://doi.org/10.1103/PhysRevD.41.3720}{\emph{Phys. Rev.
  D} {\bfseries 41} (1990) 3720}.

\bibitem{Auclair2022}
P.~Auclair and C.~Ringeval, \emph{Slow-roll inflation at {N3LO}},
  \href{https://doi.org/10.1103/physrevd.106.063512}{\emph{Phys. Rev. D 106,
  063512 (2022)} {\bfseries 106} (2022) 063512}
  [\href{https://arxiv.org/abs/2205.12608}{{\ttfamily 2205.12608}}].

\bibitem{Bianchi:2024qyp}
E.~Bianchi and M.~Gamonal, \emph{Primordial power spectrum at {N3LO} in
  effective theories of inflation},
  \href{https://doi.org/10.1103/PhysRevD.110.104032}{\emph{Phys. Rev. D}
  {\bfseries 110} (2024) 104032}
  [\href{https://arxiv.org/abs/2405.03157}{{\ttfamily 2405.03157}}].

\bibitem{Maeda:1988ab}
K.-i.~Maeda, \emph{{Towards the Einstein-Hilbert Action via Conformal
  Transformation}}, \href{https://doi.org/10.1103/PhysRevD.39.3159}{\emph{Phys.
  Rev. D} {\bfseries 39} (1989) 3159}.

\bibitem{DeFelice2010}
A.~De~Felice and S.~Tsujikawa, \emph{f(R) theories},
  \href{https://doi.org/10.12942/lrr-2010-3}{\emph{Living Rev. Rel. 13: 3,
  2010} {\bfseries 13} (2010) }
  [\href{https://arxiv.org/abs/1002.4928}{{\ttfamily 1002.4928}}].

\bibitem{Ketov:2025nkr}
S.V.~Ketov, \emph{{On Legacy of Starobinsky Inflation}},  1, 2025
  [\href{https://arxiv.org/abs/2501.06451}{{\ttfamily 2501.06451}}].

\bibitem{LIGOScientific:2016aoc}
{\scshape LIGO Scientific and Virgo} collaboration, \emph{{Observation of
  Gravitational Waves from a Binary Black Hole Merger}},
  \href{https://doi.org/10.1103/PhysRevLett.116.061102}{\emph{Phys. Rev. Lett.}
  {\bfseries 116} (2016) 061102}
  [\href{https://arxiv.org/abs/1602.03837}{{\ttfamily 1602.03837}}].

\bibitem{LIGOScientific:2017zic}
{\scshape LIGO Scientific, Virgo, Fermi-GBM, INTEGRAL} collaboration,
  \emph{{Gravitational Waves and Gamma-rays from a Binary Neutron Star Merger:
  GW170817 and GRB 170817A}},
  \href{https://doi.org/10.3847/2041-8213/aa920c}{\emph{Astrophys. J. Lett.}
  {\bfseries 848} (2017) L13}
  [\href{https://arxiv.org/abs/1710.05834}{{\ttfamily 1710.05834}}].

\bibitem{LIGOScientific:2020tif}
{\scshape LIGO Scientific and Virgo} collaboration, \emph{{Tests of General
  Relativity with Binary Black Holes from the second LIGO-Virgo
  Gravitational-Wave Transient Catalog}},
  \href{https://doi.org/10.1103/PhysRevD.103.122002}{\emph{Phys. Rev. D}
  {\bfseries 103} (2021) 122002}
  [\href{https://arxiv.org/abs/2010.14529}{{\ttfamily 2010.14529}}].

\bibitem{Vilenkin1985}
A.~Vilenkin, \emph{{Classical and Quantum Cosmology of the Starobinsky
  Inflationary Model}},
  \href{https://doi.org/10.1103/PhysRevD.32.2511}{\emph{Phys. Rev. D}
  {\bfseries 32} (1985) 2511}.

\bibitem{Delhom:2022vae}
A.~Delhom, A.~Jim\'enez-Cano and F.J.~Maldonado~Torralba, \emph{{Instabilities
  in Field Theory: A Primer with Applications in Modified Gravity}},
  SpringerBriefs in Physics, Springer, 7, 2022,
  \href{https://doi.org/10.1007/978-3-031-40433-7}{DOI}
  [\href{https://arxiv.org/abs/2207.13431}{{\ttfamily 2207.13431}}].

\bibitem{Crisostomi:2017ugk}
M.~Crisostomi, K.~Noui, C.~Charmousis and D.~Langlois, \emph{{Beyond Lovelock
  gravity: Higher derivative metric theories}},
  \href{https://doi.org/10.1103/PhysRevD.97.044034}{\emph{Phys. Rev. D}
  {\bfseries 97} (2018) 044034}
  [\href{https://arxiv.org/abs/1710.04531}{{\ttfamily 1710.04531}}].

\bibitem{Dirac:1938}
P.A.M.~Dirac, \emph{{Classical theory of radiating electrons}},
  \href{https://doi.org/10.1098/rspa.1938.0124}{\emph{Proc. Roy. Soc. Lond. A}
  {\bfseries 167} (1938) 148}.

\bibitem{Landau:1975}
L.D.~Landau and E.M.~Lifschits, \emph{{The Classical Theory of Fields}},
  vol.~Volume 2 of \emph{Course of Theoretical Physics}, Pergamon Press, Oxford
  (1975).

\bibitem{Cole:2017zca}
J.M.~Cole et~al., \emph{{Experimental evidence of radiation reaction in the
  collision of a high-intensity laser pulse with a laser-wakefield accelerated
  electron beam}}, \href{https://doi.org/10.1103/PhysRevX.8.011020}{\emph{Phys.
  Rev. X} {\bfseries 8} (2018) 011020}
  [\href{https://arxiv.org/abs/1707.06821}{{\ttfamily 1707.06821}}].

\bibitem{Nielsen:2021ppf}
{\scshape CERN NA63} collaboration, \emph{{Experimental verification of the
  Landau\textendash{}Lifshitz equation}},
  \href{https://doi.org/10.1088/1367-2630/ac1554}{\emph{New J. Phys.}
  {\bfseries 23} (2021) 085001}.

\bibitem{Frob:2013ht}
M.B.~Fr{\"o}b, D.B.~Papadopoulos, A.~Roura and E.~Verdaguer,
  \emph{{Nonperturbative semiclassical stability of de Sitter spacetime for
  small metric deviations}},
  \href{https://doi.org/10.1103/PhysRevD.87.064019}{\emph{Phys. Rev. D}
  {\bfseries 87} (2013) 064019}
  [\href{https://arxiv.org/abs/1301.5261}{{\ttfamily 1301.5261}}].

\bibitem{Ruzmaikina1969}
T.V.~Ruzma{\v{i}}kina and A.A.~Ruzma{\v{i}}kin, \emph{Quadratic corrections to
  the lagrangian density of the gravitational field and the singularity},
  {\href{http://jetp.ras.ru/cgi-bin/e/index/e/30/2/p372?a=list}{\emph{J. Exp.
  Theor. Phys.} {\bfseries 30} (1969) 372}}.

\bibitem{Starobinsky:1983zz}
A.A.~Starobinsky, \emph{{The Perturbation Spectrum Evolving from a Nonsingular
  Initially De-Sitter Cosmology and the Microwave Background Anisotropy}},
  {\emph{Sov. Astron. Lett.} {\bfseries 9} (1983) 302}.

\bibitem{Capozziello1997}
S.~Capozziello, R.~de~Ritis and A.A.~Marino, \emph{Some aspects of the
  cosmological conformal equivalence between 'Jordan frame' and 'Einstein
  frame'}, \href{https://doi.org/10.1088/0264-9381/14/12/010}{\emph{Class.
  Quant. Grav.} {\bfseries 14} (1997) 3243}
  [\href{https://arxiv.org/abs/gr-qc/9612053}{{\ttfamily gr-qc/9612053}}].

\bibitem{Faraoni1999}
V.~Faraoni, E.~Gunzig and P.~Nardone, \emph{Conformal transformations in
  classical gravitational theories and in cosmology}, {\emph{Fund. Cosmic
  Phys.} {\bfseries 20} (1999) 121}
  [\href{https://arxiv.org/abs/gr-qc/9811047}{{\ttfamily gr-qc/9811047}}].

\bibitem{Karam2017}
A.~Karam, T.~Pappas and K.~Tamvakis, \emph{Frame-dependence of higher-order
  inflationary observables in scalar-tensor theories},
  \href{https://doi.org/10.1103/physrevd.96.064036}{\emph{Phys. Rev. D 96,
  064036 (2017)} {\bfseries 96} (2017) 064036}
  [\href{https://arxiv.org/abs/1707.00984}{{\ttfamily 1707.00984}}].

\bibitem{Ketov:2024klm}
S.V.~Ketov, \emph{{Starobinsky inflation and swampland conjectures}},
  \href{https://doi.org/10.1007/s11182-024-03318-7}{\emph{Russ. Phys. J.}
  {\bfseries 67} (2024) 1819}
  [\href{https://arxiv.org/abs/2406.06923}{{\ttfamily 2406.06923}}].

\bibitem{Toyama:2024ugg}
S.~Toyama and S.V.~Ketov, \emph{{Starobinsky inflation beyond the leading
  order}}, \href{https://doi.org/10.1103/PhysRevD.110.063552}{\emph{Phys. Rev.
  D} {\bfseries 110} (2024) 063552}
  [\href{https://arxiv.org/abs/2407.21349}{{\ttfamily 2407.21349}}].

\bibitem{Appleby:2009uf}
S.A.~Appleby, R.A.~Battye and A.A.~Starobinsky, \emph{{Curing singularities in
  cosmological evolution of F(R) gravity}},
  \href{https://doi.org/10.1088/1475-7516/2010/06/005}{\emph{JCAP} {\bfseries
  06} (2010) 005} [\href{https://arxiv.org/abs/0909.1737}{{\ttfamily
  0909.1737}}].

\bibitem{Kannike:2015apa}
K.~Kannike, G.~H{\"u}tsi, L.~Pizza, A.~Racioppi, M.~Raidal, A.~Salvio et~al.,
  \emph{{Dynamically Induced Planck Scale and Inflation}},
  \href{https://doi.org/10.1007/JHEP05(2015)065}{\emph{JHEP} {\bfseries 05}
  (2015) 065} [\href{https://arxiv.org/abs/1502.01334}{{\ttfamily
  1502.01334}}].

\bibitem{vandeBruck:2016leo}
C.~van~de Bruck, P.~Dunsby and L.E.~Paduraru, \emph{{Reheating and preheating
  in the simplest extension of Starobinsky inflation}},
  \href{https://doi.org/10.1142/S0218271817501528}{\emph{Int. J. Mod. Phys. D}
  {\bfseries 26} (2017) 1750152}
  [\href{https://arxiv.org/abs/1606.04346}{{\ttfamily 1606.04346}}].

\bibitem{Nandi:2019xlj}
D.~Nandi, \emph{{Note on stability in conformally connected frames}},
  \href{https://doi.org/10.1103/PhysRevD.99.103532}{\emph{Phys. Rev. D}
  {\bfseries 99} (2019) 103532}
  [\href{https://arxiv.org/abs/1904.00153}{{\ttfamily 1904.00153}}].

\bibitem{Nandi:2019xve}
D.~Nandi and P.~Saha, \emph{{Einstein or Jordan: seeking answers from the
  reheating constraints}},  [\href{https://arxiv.org/abs/1907.10295}{{\ttfamily
  1907.10295}}].

\bibitem{Dorsch:2024nan}
G.C.~Dorsch, L.~Miranda and N.~Yokomizo, \emph{{Gravitational reheating in
  Starobinsky inflation}},
  \href{https://doi.org/10.1088/1475-7516/2024/11/050}{\emph{JCAP} {\bfseries
  11} (2024) 050} [\href{https://arxiv.org/abs/2406.04161}{{\ttfamily
  2406.04161}}].

\bibitem{Bianchi:2024jmn}
E.~Bianchi and M.~Gamonal, \emph{{Squeezed vacua and primordial features in
  effective theories of inflation at N2LO}},
  \href{https://doi.org/10.1103/hjlj-sv2t}{\emph{Phys. Rev. D} {\bfseries 111}
  (2025) 124024} [\href{https://arxiv.org/abs/2410.11812}{{\ttfamily
  2410.11812}}].

\bibitem{Stewart2001}
E.D.~Stewart and J.-O.~Gong, \emph{The density perturbation power spectrum to
  second-order corrections in the slow-roll expansion},
  \href{https://doi.org/10.1016/s0370-2693(01)00616-5}{\emph{Phys.Lett.B510:1-9,2001}
  {\bfseries 510} (2001) 1}
  [\href{https://arxiv.org/abs/astro-ph/0101225}{{\ttfamily
  astro-ph/0101225}}].

\bibitem{Ballardini:2024irx}
M.~Ballardini, A.~Davoli and S.S.~Sirletti, \emph{{Third-order corrections to
  the slow-roll expansion: Calculation and constraints with Planck, ACT, SPT,
  and BICEP/Keck}},
  \href{https://doi.org/10.1016/j.dark.2025.101813}{\emph{Phys. Dark Univ.}
  {\bfseries 47} (2025) 101813}
  [\href{https://arxiv.org/abs/2408.05210}{{\ttfamily 2408.05210}}].

\bibitem{Zharov:2025evb}
D.S.~Zharov, O.O.~Sobol and S.I.~Vilchinskii, \emph{{Reheating ACTs on
  Starobinsky and Higgs inflation}},
  [\href{https://arxiv.org/abs/2505.01129}{{\ttfamily 2505.01129}}].

\bibitem{ZenodoR2W2}
M.~Gamonal and E.~Bianchi, \emph{{R2W2-PowerSpectrum}},  2025.
\newblock
  \href{https://doi.org/10.5281/zenodo.16954788}{10.5281/zenodo.16954788}.

\bibitem{Racioppi:2021jai}
A.~Racioppi and M.~Vasar, \emph{{On the number of e-folds in the Jordan and
  Einstein frames}},
  \href{https://doi.org/10.1140/epjp/s13360-022-02853-x}{\emph{Eur. Phys. J.
  Plus} {\bfseries 137} (2022) 637}
  [\href{https://arxiv.org/abs/2111.09677}{{\ttfamily 2111.09677}}].

\bibitem{Euclid:2023shr}
{\scshape Euclid} collaboration, \emph{{Euclid: The search for primordial
  features}}, \href{https://doi.org/10.1051/0004-6361/202348162}{\emph{Astron.
  Astrophys.} {\bfseries 683} (2024) A220}
  [\href{https://arxiv.org/abs/2309.17287}{{\ttfamily 2309.17287}}].

\bibitem{LiteBIRDCollaboration2023}
{LiteBIRD Collaboration}, U.~Fuskeland, J.~Aumont, R.~Aurlien et~al.,
  \emph{Tensor-to-scalar ratio forecasts for extended litebird frequency
  configurations},
  \href{https://doi.org/10.1051/0004-6361/202346155}{\emph{A\&A 676, A42
  (2023)} {\bfseries 676} (2023) A42}
  [\href{https://arxiv.org/abs/2302.05228}{{\ttfamily 2302.05228}}].

\bibitem{Martin:2006rs}
J.~Martin and C.~Ringeval, \emph{{Inflation after WMAP3: Confronting the
  Slow-Roll and Exact Power Spectra to CMB Data}},
  \href{https://doi.org/10.1088/1475-7516/2006/08/009}{\emph{JCAP} {\bfseries
  08} (2006) 009} [\href{https://arxiv.org/abs/astro-ph/0605367}{{\ttfamily
  astro-ph/0605367}}].

\bibitem{Martin:2010kz}
J.~Martin and C.~Ringeval, \emph{{First CMB Constraints on the Inflationary
  Reheating Temperature}},
  \href{https://doi.org/10.1103/PhysRevD.82.023511}{\emph{Phys. Rev. D}
  {\bfseries 82} (2010) 023511}
  [\href{https://arxiv.org/abs/1004.5525}{{\ttfamily 1004.5525}}].

\bibitem{Martin:2014nya}
J.~Martin, C.~Ringeval and V.~Vennin, \emph{{Observing Inflationary
  Reheating}},
  \href{https://doi.org/10.1103/PhysRevLett.114.081303}{\emph{Phys. Rev. Lett.}
  {\bfseries 114} (2015) 081303}
  [\href{https://arxiv.org/abs/1410.7958}{{\ttfamily 1410.7958}}].

\bibitem{Drees:2025ngb}
M.~Drees and Y.~Xu, \emph{{Refined predictions for Starobinsky inflation and
  post-inflationary constraints in light of ACT}},
  \href{https://doi.org/10.1016/j.physletb.2025.139612}{\emph{Phys. Lett. B}
  {\bfseries 867} (2025) 139612}
  [\href{https://arxiv.org/abs/2504.20757}{{\ttfamily 2504.20757}}].

\bibitem{Yin:2025rrs}
W.~Yin, \emph{{Higgs-like inflation ACTivated mass}},
  \href{https://doi.org/10.1088/1475-7516/2025/09/062}{\emph{JCAP} {\bfseries
  09} (2025) 062} [\href{https://arxiv.org/abs/2505.03004}{{\ttfamily
  2505.03004}}].

\bibitem{Yogesh:2025wak}
Yogesh, A.~Mohammadi, Q.~Wu and T.~Zhu, \emph{{Starobinsky like inflation and
  EGB Gravity in the light of ACT}},
  \href{https://doi.org/10.1088/1475-7516/2025/10/010}{\emph{JCAP} {\bfseries
  10} (2025) 010} [\href{https://arxiv.org/abs/2505.05363}{{\ttfamily
  2505.05363}}].

\bibitem{Addazi:2025qra}
A.~Addazi, Y.~Aldabergenov and S.V.~Ketov, \emph{{Curvature corrections to
  Starobinsky inflation can explain the ACT results}},
  \href{https://doi.org/10.1016/j.physletb.2025.139883}{\emph{Phys. Lett. B}
  {\bfseries 869} (2025) 139883}
  [\href{https://arxiv.org/abs/2505.10305}{{\ttfamily 2505.10305}}].

\bibitem{Choudhury:2025vso}
S.~Choudhury, G.~Bauyrzhan, S.K.~Singh and K.~Yerzhanov, \emph{{What new
  physics can we extract from inflation using the ACT DR6 and DESI DR2
  Observations?}},  [\href{https://arxiv.org/abs/2506.15407}{{\ttfamily
  2506.15407}}].

\bibitem{Wolf:2025ecy}
W.J.~Wolf, \emph{{Inflationary attractors and radiative corrections in light of
  ACT}},  [\href{https://arxiv.org/abs/2506.12436}{{\ttfamily 2506.12436}}].

\bibitem{Frolovsky:2025iao}
D.~Frolovsky and S.V.~Ketov, \emph{{Are single-field models of inflation and
  PBH production ruled out by ACT observations?}},
  [\href{https://arxiv.org/abs/2505.17514}{{\ttfamily 2505.17514}}].

\bibitem{Chluba:2015bqa}
J.~Chluba, J.~Hamann and S.P.~Patil, \emph{{Features and New Physical Scales in
  Primordial Observables: Theory and Observation}},
  \href{https://doi.org/10.1142/S0218271815300232}{\emph{Int. J. Mod. Phys. D}
  {\bfseries 24} (2015) 1530023}
  [\href{https://arxiv.org/abs/1505.01834}{{\ttfamily 1505.01834}}].

\bibitem{Borissova:2022clg}
J.N.~Borissova and B.~Dittrich, \emph{{Towards effective actions for the
  continuum limit of spin foams}},
  \href{https://doi.org/10.1088/1361-6382/accbfb}{\emph{Class. Quant. Grav.}
  {\bfseries 40} (2023) 105006}
  [\href{https://arxiv.org/abs/2207.03307}{{\ttfamily 2207.03307}}].

\bibitem{Dittrich:2023rcr}
B.~Dittrich and J.~Padua-Arg\"uelles, \emph{{Lorentzian Quantum Cosmology from
  Effective Spin Foams}},
  \href{https://doi.org/10.3390/universe10070296}{\emph{Universe} {\bfseries
  10} (2024) 296} [\href{https://arxiv.org/abs/2306.06012}{{\ttfamily
  2306.06012}}].

\bibitem{Gielen:2025jcb}
S.~Gielen and L.~Mickel, \emph{{Cosmological scalar perturbations for a metric
  reconstructed from group field theory}},
  [\href{https://arxiv.org/abs/2505.07951}{{\ttfamily 2505.07951}}].

\bibitem{Oriti:2025lwx}
X.~Pang and D.~Oriti, \emph{{Late-time cosmic acceleration from quantum
  gravity}}, \href{https://doi.org/10.1088/1361-6382/adecdb}{\emph{Class.
  Quant. Grav.} {\bfseries 42} (2025) 155003}
  [\href{https://arxiv.org/abs/2502.12419}{{\ttfamily 2502.12419}}].

\bibitem{MenaMarugan:2024vyy}
G.A.~Mena~Marug\'an, A.~Vicente-Becerril and J.~Y\'ebana~Carrilero,
  \emph{{Comparing Analytic and Numerical Studies of Tensor Perturbations in
  Loop Quantum Cosmology}},
  \href{https://doi.org/10.3390/universe10090365}{\emph{Universe} {\bfseries
  10} (2024) 365} [\href{https://arxiv.org/abs/2409.18302}{{\ttfamily
  2409.18302}}].

\bibitem{MenaMarugan:2024zcv}
G.A.~Mena~Marugan, A.~Vicente-Becerril and J.~Yebana~Carrilero, \emph{{Analytic
  and numerical study of scalar perturbations in loop quantum cosmology}},
  \href{https://doi.org/10.1103/PhysRevD.110.043508}{\emph{Phys. Rev. D}
  {\bfseries 110} (2024) 043508}
  [\href{https://arxiv.org/abs/2404.04595}{{\ttfamily 2404.04595}}].

\end{thebibliography}

%%%%%%%

\providecommand{\href}[2]{#2}\begingroup\raggedright\endgroup

\end{document}